\documentclass[prb,superscriptaddress,showpacs,twocolumn]{revtex4}

\usepackage{graphicx}

\usepackage{amssymb,amsmath}
\usepackage[nooneline,raggedright]{subfigure}

\usepackage[dvips]{color}

\newcommand{\eref}[1]{Eq.~(\ref{#1})}
\newcommand{\fref}[1]{Fig.~\ref{#1}}
\newcommand{\sref}[1]{Section~\ref{#1}}
\newcommand{\tref}[1]{Tab.~\ref{#1}}

\newcommand{\im}{%
           \imath}
\newcommand{\bra}[1]{\ensuremath{\langle #1|}}
\newcommand{\ket}[1]{\ensuremath{|#1\rangle}}

\newcommand{\ppr}{%
        ^{\prime\prime}}

\def\t2g{\ensuremath{t_{2g}}}
\def\a1g{\ensuremath{a_{1g}}}

\newcommand{\vek}[1]{%
        \hbox{\textbf #1}}

\newcommand{\pr}{%
        ^\prime}

\def\etal{{\it et~al.}}

\def\XXint#1#2#3{{\setbox0=\hbox{$#1{#2#3}{\int}$}
\vcenter{\hbox{$#2#3$}}\kern-.5\wd0}}

\newcommand{\oo}[1]{\frac{1}{#1}}

\newcommand{\fesb}{FeSb$_2$}
\newcommand{\feas}{FeAs$_2$}

\newcounter{fnnumber}

\begin{document}

\title{Thermopower of correlated semiconductors~: application to \feas\ and \fesb}

\author{Jan M. Tomczak}
\affiliation{Department of Physics and Astronomy, Rutgers University, Piscataway, New Jersey 08854, USA}
\author{K. Haule}
\affiliation{Department of Physics and Astronomy, Rutgers University, Piscataway, New Jersey 08854, USA}
\author{T. Miyake}
\affiliation{Nanosystem Research Institute, AIST, Tsukuba 305-8568, Japan}
\affiliation{Japan Science and Technology Agency, CREST, Kawaguchi 332-0012, Japan}
\author{A. Georges}
\affiliation{Centre de Physique Th{\'e}orique, Ecole Polytechnique, CNRS, 91128 Palaiseau Cedex, France}
\author{G. Kotliar}
\affiliation{Department of Physics and Astronomy, Rutgers University, Piscataway, New Jersey 08854, USA}

\begin{abstract}
We investigate the effect of electronic correlations onto the thermoelectricity of semi-conductors and insulators.
Appealing to model considerations, we study various many-body renormalizations that enter the thermoelectric response.
We find that, contrary to the case of correlated metals, correlation effects do not {\it per se} enhance the Seebeck coefficient
or the figure of merit, for the former of which we give an upper bound in the limit of vanishing vertex corrections. 
For two materials of current interest, \feas\ and \fesb, we compute the electronic structure and thermopower.
We find \feas\ to be well described within density functional theory, and the therefrom deduced Seebeck coefficient to
be in quantitative agreement with experiment. The capturing of the insulating ground state of \fesb, however, requires the inclusion of many-body effects, in which we succeed by applying the GW approximation. Yet, while we get qualitative agreement for the thermopower of \fesb\ at intermediate temperatures, the tremendously large Seebeck coefficient at low temperatures is found to  violate our upper bound, suggesting the presence of decisive (e.g.\ phonon mediated) vertex corrections. 
\end{abstract}

\maketitle

\section{Introduction}

While recent efforts to design materials with enhanced thermoelectric
properties were mainly focused on reducing the lattice contributions
to the thermal conductivity by super- or nano
structures,\cite{Snyder2008} interest in the potential merits of
electronic correlation effects was revived by the discovery of large
Seebeck coefficients in transition metal compounds, such as
FeSi,\cite{JPSJ.76.093601}
Na$_x$CoO$_2$,\cite{PhysRevB.56.R12685} and
\fesb\cite{0295-5075-80-1-17008}, with the latter displaying an
astonishing response of up to $S=-45$~mV/K at
12K~\cite{0295-5075-80-1-17008}.  Indeed, on a model
level,\cite{PhysRevLett.67.3724,PhysRevLett.80.4775,PhysRevB.65.075102,grenzebach_2006}
as well as for realistic
compounds,\cite{oudovenko:035120,haule_thermo} correlation effects
were shown to enhance the Seebeck coefficient in metals and 
transition metal oxides~\cite{Held_thermo,arita:115121}.

In this work we address the thermoelectric response of correlated
semiconductors and insulators.  Our aim is to set up the general 
formalism for discussing the thermoelectric response, with the
ultimate goal to understand the origin of the large thermoelectricity
observed in correlated semiconductors and to search for high performance
thermoelectrics in this class of materials.  This work parallels the
analysis done for correlated metals in
Refs.~\cite{PhysRevLett.80.4775,haule_thermo}.

In particular, we investigate whether the electronic structure and
correlation effects alone can account for the very different
magnitudes in the Seebeck coefficient of the two iso-structural and
iso-electronic compounds \fesb\ and \feas.  In the case of \fesb, it
has indeed been conjectured that electronic correlations are at the
origin of the huge thermoelectric
response~\cite{0295-5075-80-1-17008,sun:153308,APEX.2.091102,sun_dalton}.

The setup of the paper is the following. First we describe the FeX$_2$
materials. Then we shall first extent general text-book considerations
for the Seebeck coefficient to include the important aspect of carrier
selective electronic renormalizations, as well as properties beyond
the picture of coherent band-structures.  We use this framework to
make general arguments on how to obtain high values for the figure of
merit $ZT$.  Given the sizes of the charge gaps, effective masses, and
other parameters, these considerations allow us to put constraints on
the possible regimes of materials of interest.
In \sref{estruc} we apply realistic electronic structure tools, and
show that for iron arsenide, \feas, both the electronic structure, as
well as the thermoelectric response can be understood as that of a
conventional semiconductor and described quantitatively by {\it ab
  initio} band-structure methods.  For the iron antimonide on the
other hand, standard density functional theory
(DFT)\cite{RevModPhys.71.1253} based methods are known to be
insufficient to account for the electronic
structure\cite{luko_fesb2,Madsen_fesb2}.  To show that correlation
effects play an important role, we employ a hybrid functional
approach\cite{becke:1372}, as well as Hedin's GW
approximation\cite{hedin}, with the latter yielding results in good
agreement with the experimental charge gap.  As to the thermoelectric
response, however, we find qualitative agreement with experiment only
at intermediate temperatures (35-70K).  Our analysis shows that the
low temperature Seebeck coefficient of \fesb\ is incompatible with a local electronic picture, suggesting the importance of
vertex corrections and non-local self-energy effects (that we
neglect), or the presence of a substantial phonon drag
effect,\cite{PhysRev.96.1163} as is e.g.\ found in the classical
example of p-type Germanium~\cite{PhysRev.94.1134}.

\section{The materials}

Despite the structural similarity of \fesb, and \feas, experimental findings point to markedly different properties, heralding a varying importance of correlation, and, potentially, electron-phonon effects.

\feas\ is an insulator with a gap of 0.2-0.22~eV~\cite{Fan1972136,APEX.2.091102}, as obtained from the activation behavior in the resistivity at temperatures of 200K and higher.
Below 200K, the influence of impurities is pivotal\cite{Fan1972136,APEX.2.091102}~: the resistivity has a metallic slope before resuming, below 30K and down to 10K, an activation
law with an energy of 0.01eV. Further, below 10K, the resistivity exhibits activation with 6K (0.5meV).\cite{APEX.2.091102}
The Hall coefficient is negative for all temperatures~\cite{Fan1972136}. Congruently, the Seebeck coefficient is negative as well~: From its room temperature value
$-200\mu$V/K~\cite{Fan1972136}, it grows in magnitude upon cooling, to reach $-7$mV/K at 12K,\cite{APEX.2.091102} before it vanishes towards zero temperature.\cite{APEX.2.091102}

In the case of \fesb, optical spectroscopy finds a small gap of 432K (37meV) at low temperatures,\cite{perucci_optics} 
but witnesses the development of a Drude-like peak at 70K and above. The concomitant transfer of spectral weight is found 
to extend over an energy range as high as 1eV~\cite{perucci_optics}, i.e.\ a scale that is much larger than the initial gap, a
common harbinger of correlation effects~\cite{PhysRevB.54.8452}.
The resistivity of \fesb, on the other hand, has three distinct temperature regimes that exhibit activated behavior~:
In the range of 50--100K the activation energy corresponds to a gap of 300K (26meV)~\cite{PhysRevB.67.155205,0295-5075-80-1-17008}.
From 20K down to 10K, the resistivity shows a shoulder-like behavior with an activation energy of $\Delta/2=3$meV, 
while below 5K, extrinsic impurities are believed to be at the origin of a weakly temperature dependent resistivity following an activation
behavior with 0.04-0.09meV.\cite{0295-5075-80-1-17008}
The resistivity is anisotropic~\cite{Fan1972136,PhysRevB.67.155205,0295-5075-80-1-17008}, and some experiments find metallic transport
behavior ($d\rho/dT>0$)  for selected directions above 40K.~\cite{PhysRevB.67.155205,PhysRevB.74.195130}
This anisotropy is also seen in Hall measurements.\cite{Fan1972136}   
As a matter of fact, the Hall coefficient even changes sign for some polarizations (at 100K,\cite{Fan1972136} or 40K\cite{hu:182108}),
with predominant electron character (n-type) below these temperatures.
The Seebeck coefficient at 300K is found to be $15-40\mu$V/K, with the sign depending on the polarization.~\cite{0295-5075-80-1-17008,bentien:205105} 
Upon on lowering the temperature, the Seebeck coefficient passes a local maximum ($\partial^2 S/\partial T^2 >0$) at around 40K, before 
turning towards very large negative values, reaching, depending on the polarization and the sample, a global extrema of up to $-45$mV/K at slightly above 10K.~\cite{0295-5075-80-1-17008}
Below this temperature, the coefficient drops sharply in magnitude and practically vanishes at 5K and below.
Interestingly, the largest thermopower is thus found in the temperature range where the resistivity has the shoulder-like behavior.
Noteworthy, this regime is concomitant with the appearance of a prominent feature in the ``electronic'' specific heat~\cite{APEX.2.091102}. No such feature is found for the arsenide\footnote{Frank Steglich  and Peijie Sun -- private communication}. \setcounter{fnnumber}{\thefootnote}
However, the unlocking of spins in \fesb\ becomes appreciable only beyond this regime at around 150K, where the entropy reaches $R\log2$, owing to a second and larger hump in the specific heat,
and in congruity with the susceptibility\cite{koyama:073203,PhysRevB.67.155205}. 
Indeed \fesb\ becomes paramagnetic above  100K~\cite{PhysRevB.67.155205}, and a Curie-like downturn appears at temperatures above 350K~\cite{koyama:073203}, whereas the susceptibility of \feas\ is flat up to 350K\cite{APEX.2.091102}.
That the low temperature feature in the specific heat of \fesb\ has no spin signature might indicate that its contribution to the entropy is
associated with either the charge degrees of freedom or an electron-phonon effect. 
The importance of electron-electron effects in \fesb\ is further highlighted by the fact that
various properties are very sensitive with respect to changes in the carrier density. Doping the system with electrons, e.g.\ FeSb$_{2-x}$Sn$_x$~\cite{bentien:205105},  or holes, e.g.\ FeSb$_{2-x}$Te$_x$\cite{hu:064510,sun:033710}, instantly metalizes the compound, sometimes generates a Curie law at low temperatures~\cite{hu:064510} 
and reduces the Seebeck coefficient~\cite{bentien:205105,sun:033710}. 
Thermoelectric properties of FeSb$_{2-x}$Te$_x$
are indeed that of a correlated metal, i.e.\ the low temperature Seebeck coefficient is linear in T, with an enhancement factor of 15 via the effective mass~\cite{sun_dalton}.

Despite these indications for correlation effects, some experimental findings for the antimonide are quantitatively reproducible by conventional band-structure methods~: Volume and bulk-modulus~\cite{PhysRevB.72.045103} are very well
captured within the generalized gradient approximation (GGA) of DFT~\cite{0953-8984-21-18-185403}.
Also, the finding of small electron and hole pockets in \fesb\cite{luko_fesb2,Madsen_fesb2} (see also below) within band-structure methods could simply be attributed to the
well documented underestimation of charge gaps within DFT. 
Moreover, a calculation within the local density approximation (LDA) with a Hartree-like Coulomb interaction (LDA+U) suggested that \fesb, while being paramagnetic, could be close to a ferromagnetic instability~\cite{luko_fesb2}.
Weak ferromagnetism was then indeed found in Fe$_{1-x}$Co$_x$Sb$_2$~\cite{hu:224422}.

\section{Transport formluae, general \& model considerations}
Within the Kubo formalism the Seebeck coefficient -- that relates the
gradients of temperature and electrical field -- is given by (see
e.g. Ref.~\onlinecite{oudovenko:035120,haule_thermo})
\begin{eqnarray}
S&=&-\frac{k_B}{\left|e\right|}  \frac{A_1}{A_0}
\label{eqS}
\end{eqnarray}
where the current-current (current-heat current) correlation function $A_0$ ($A_1$) is given by
\begin{eqnarray}
A_n&=&\int d\omega \beta^n(\omega-\mu)^n \left(-\frac{\partial f_\mu}{\partial \omega}  \right) \Xi(\omega)
\label{A01}
\end{eqnarray}
Here, $f_\mu$ is the Fermi function, $\mu$ is the Fermi level, and
$\Xi$ is the transport kernel. If vertex corrections are neglected,
the transport kernel can be expressed (in matrix notation) as 
\begin{eqnarray}
\Xi(\omega)&=&\sum_{\vek{k}} \textrm{Tr}\left[ v(\vek{k})A(\vek{k},\omega)v(\vek{k})A(\vek{k},\omega)\right]
\end{eqnarray}
with the Fermi velocity
$v_{ij}(\vek{k})=-\frac{ie}{m}\bra{\Psi_{\vek{k}i}}\nabla\ket{\psi_{\vek{k}j}}$, and
the spectral function $A_{ij}(\vek{k},\omega)$, where $\psi_{\vek{k}i}$ is a
complete set of one electron basis functions, such as Kohn-Sham orbitals.
Using transport coefficients \eref{A01}, we can further express (see
e.g.\ Ref.~\onlinecite{PhysRevLett.80.4775}) the dc conductivity, the
thermal conductivity, and the figure of merit as
\begin{eqnarray}
\sigma&=&\frac{2\pi e^2}{\hbar V}A_0\\
\kappa&=&\kappa_L+\frac{2\pi k_B^2}{\hbar V}T\left(
A_2-\frac{A_1^2}{A_0}\right)\\
ZT&=& \frac{S^2 \sigma T}{\kappa}
\end{eqnarray}
where $V$ is the unit-cell volume, and $\kappa_L$ the thermal lattice conductivity. 

The chemical potential, $\mu$, is obtained by the requirement of the charge
neutrality,
\begin{equation}
n-p-n_{D^+}=0
\label{neutral}
\end{equation}
where 
\begin{eqnarray}
\left.\begin{array}{l}
n\\
p \end{array}\right\}&=&\sum_{\vek{k}}\int_{-\infty}^\infty d\omega \left\{\begin{array}{l}
A^v(\vek{k},\omega)f_\mu(\omega)\\
A^c(\vek{k},\omega)\left[1-f_\mu(\omega)\right] \end{array}\right. 
\label{eqnp}
\end{eqnarray}
is the number of electrons (holes). Here, $A^{c,v}(k,\omega)$ is the valence/conduction spectral function. 
We also allowed for
the presence of ionized donor impurities $n_{D^+}=n_D \left[1+2e^{-\beta(E_D-\mu)} \right]^{-1}$ of concentration $n_D$ at an energy $E_D$. 

\subsection{model considerations}

Here we find it instructive to extend on the usual text-book
considerations (see e.g. Ref.~\onlinecite{ziman_ep}) and generalize to include
carrier dependent masses, renormalizations, as well as finite (yet
energy-independent) scattering amplitudes. This makes it possible to
investigate the important effects of particle-hole asymmetry, carrier
coherence, and allows us later to discuss the consistency of a purely
diffusive thermopower for a given material.

Assuming a Lorentzian line shape of the conduction (c) and valence (v) spectral functions
\begin{eqnarray}
A^{c,v}_{coh}(k,\omega)&=&\frac{Z_k^{c,v}}{\pi}\frac{\Gamma_k^{c,v}}{(\omega-\xi_k^{c,v})^2+(\Gamma_k^{c,v})^2}
\end{eqnarray}
i.e.\ we limit the discussion to the coherent part, $A_{coh}$, (of
weight $Z_k$) of the full spectrum, $A=A_{coh}+A_{incoh}$, with a quasi-particle dispersion $\xi_k$, and an
elastic scattering of amplitude $\Gamma_k$. Within this approximation,
one finds for the number of electrons%
\footnote{As is the usual custom when not knowing the incoherent part of the spectrum (as e.g.\ in slave boson techniques), we omit the $Z$-factors in the 
determination of the particle numbers.}
\begin{eqnarray}
n=\sum_k\left\{\oo{2}-\oo{\pi} Im \psi\left[\oo{2}+\frac{\beta}{2\pi}\biggl(\Gamma_k+\im(\xi_k-\mu)\biggr)\right]\right\}
\label{n}
\end{eqnarray}
where $\psi(z)$ 
is the digamma function, and $\beta=1/(k_BT)$ the inverse temperature. 
The above expression reduces to the usual $\sum_k f(\mu-\xi_k)$ 
in the coherent limit ($\Gamma=0$).
The response functions, \eref{A01}, can be also be expressed analytically (here, we restrict ourselves to $n=0,\,1$) as  
\begin{eqnarray}
\label{XXX}
A_n&=&\frac{\beta^{n+1}}{4\pi^3} \sum_k \frac{Z_k^2}{\Gamma_k} v_k^2 \biggl\{ (\xi_k-\mu)^n Re\psi\pr (x) \biggr.\\ 
&&\,\,\biggl.-\frac{\beta}{2\pi}\biggl( (\xi_k-\mu)^n  \Gamma_k Re \psi\ppr (x)+ n \Gamma^2_k Im \psi\ppr (x)\biggr) \biggr\}\nonumber
\end{eqnarray}
where the arguments of the derivatives of the digamma function are
\begin{equation}
x=\left[\oo{2}+\frac{\beta}{2\pi}\biggl(\Gamma_k-\im(\xi_k-\mu)\biggr)\right] \nonumber
\end{equation}

The first contribution in the curly brackets of \eref{XXX} is the leading term in the coherent limit ($\Gamma\rightarrow 0$). Indeed with $\beta/(2\pi^2)Re\psi\pr(1/2+i\beta x/(2\pi)=-f\pr(x)$ one recovers the Boltzmann expressions for transport coefficients,\cite{ziman_ep} in which appears an {\it ad hoc} lifetime $\tau=1/(2\Gamma)$.
In the above expression, however, the influence of {\it finite} scatterings is not restricted to the pre--factor, but since
the spectrum broadens, a wider energy range becomes activated for supplying charge carriers, and, as a result, the Fermi statistics assumes the digamma form,
and, also, higher order terms appear. Therewith, contrary to the Boltzmann description at small $\Gamma$, coherence effects do not in general cancel in the ratio $A_1/A_0$ of the Seebeck coefficient.

For illustrative purposes, we now analyze \eref{XXX} in terms of  special cases for simple quadratic dispersions~:
We consider bare bands  $\epsilon^k_{c,v}=\frac{\hbar^2k^2}{2m^0_{c,v}}$, and an interacting dispersion 
\begin{eqnarray}
\xi^k_{c,v}&=&\pm\Delta/2\pm\frac{\hbar^2k^2}{2m^*_{c,v}}
\label{eqxi}
\end{eqnarray}
where $\Delta$ is the charge gap, $m^*$ the effective mass of the carriers, and the origin of the chemical potential is chosen at the mid-gap point.
Further, we assume the Fermi velocities to be given by the group velocity%
\footnote{for a discussion on this kind of Peierls approximation see e.g. \cite{optic_prb}}%
\begin{eqnarray}
v_k&=&\frac{1}{\hbar}\partial_k\epsilon_k
\label{groupvelo}
\end{eqnarray}
and weights, $Z$, and scattering rates, $\Gamma$, that are independent of momentum.

\begin{figure*}[!t!h]
  \begin{center}
\subfigure[\,coefficient of the $1/T$ behavior in the gap.]{\scalebox{1.}{\includegraphics[angle=-90,width=.425\textwidth]{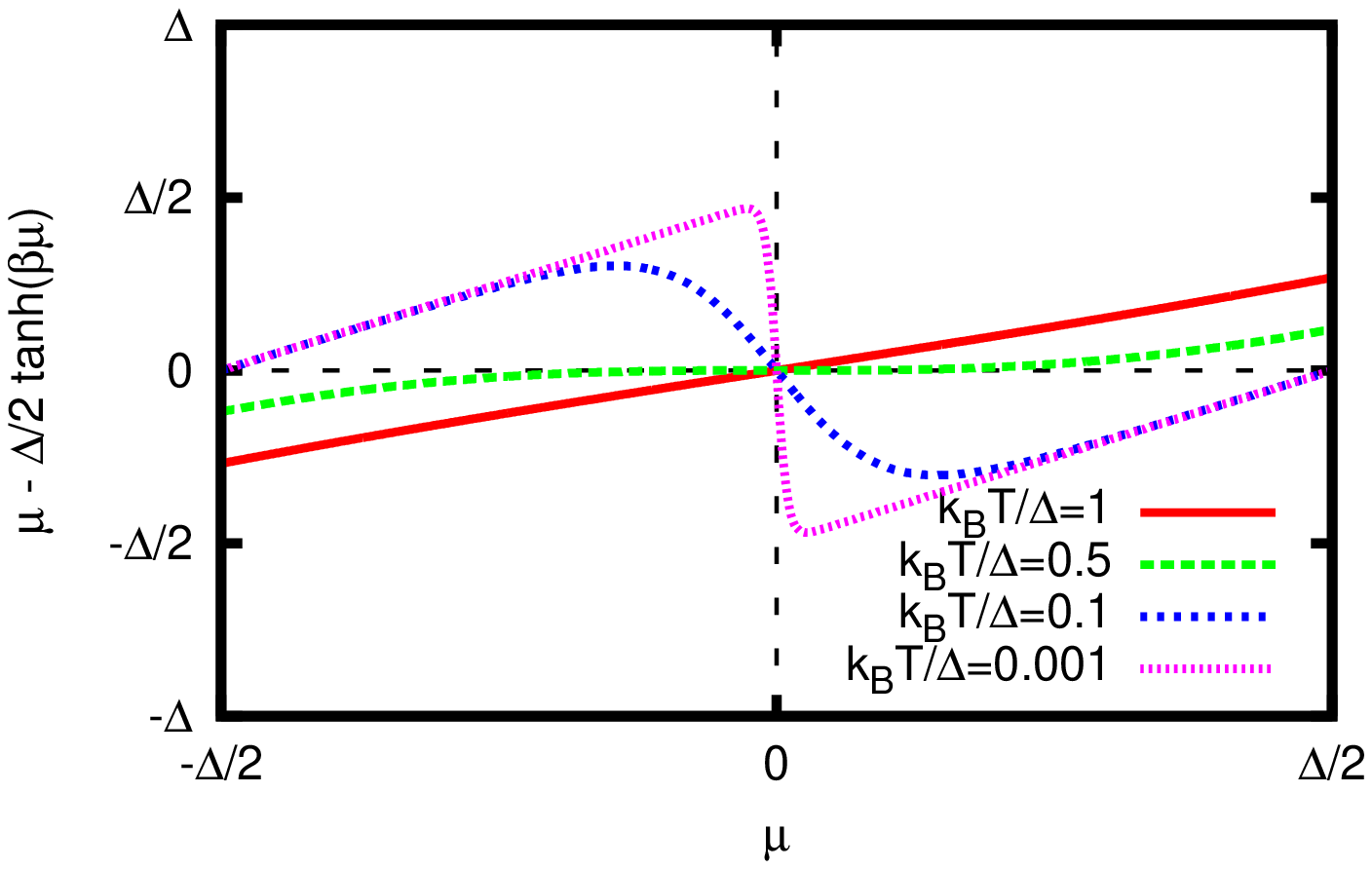}}} 
\subfigure[\,maximum value of the $1/T$ coefficient.]{\scalebox{1.}{\includegraphics[angle=-90,width=.425\textwidth]{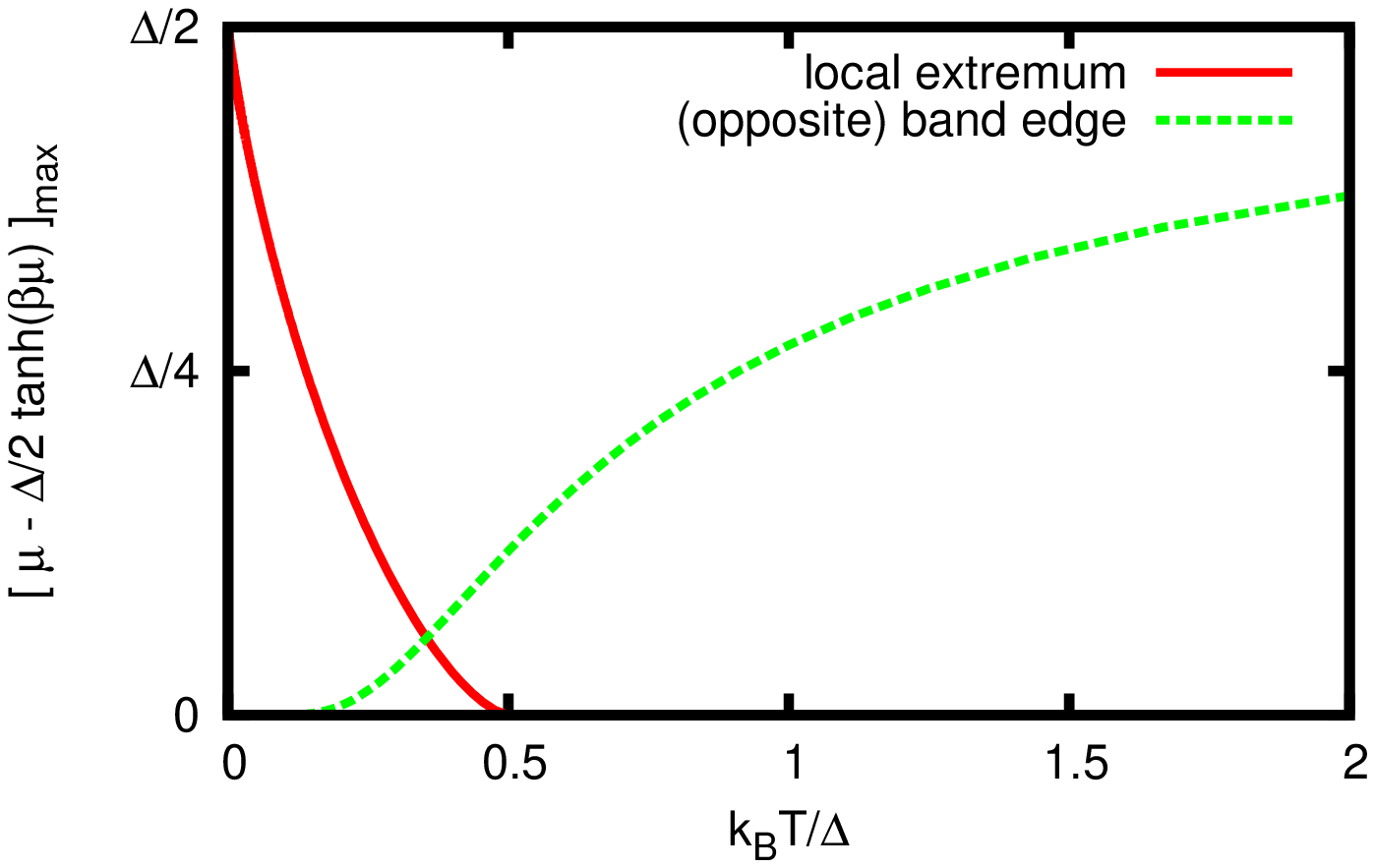}}} 
      \caption{The symmetric, coherent, large gap semiconductor. (a) coefficient of the $1/T$ low temperature behavior of the Seebeck coefficient as a function
      of the chemical potential, and for different temperatures. (b) temperature evolution of the extremal value of the $1/T$ coefficient of the thermopower.}
      \label{figCmodel}
      \end{center}
\end{figure*}

\subsubsection{the large gap coherent semi-conductor}
In the limit of a coherent system ($\Gamma\ll 1$) with a large gap ($\beta\left|\Delta/2-\mu\right|\gg1$), \eref{XXX} can be simplified to
\begin{eqnarray}
&&A_n^{c/v}=\\
&& (\pm 1)^n \frac{3\lambda}{\sqrt{2\pi^5\beta^3}}  e^{-\beta\Delta/2} \left\{ [\beta(\pm\mu-\Delta/2)]^n  - \frac{5}{2} n \right\}  \nonumber
\end{eqnarray}
where all carrier specific parameters have been gathered in
\begin{eqnarray}
\lambda_{c,v}&=&\frac{Z^2}{\Gamma}\frac{m_*^{5/2}}{m_0^2} e^{\pm \beta\mu}
\end{eqnarray}
Therewith the Seebeck coefficient becomes
\begin{eqnarray}
S&=&-\frac{k_B}{\left|e\right|}  \frac{A_1^c+A_1^v}{A_0^c+A_0^v}\\
&=& \oo{ \left| e \right| T} \left( \mu-\frac{\Delta}{2}\delta\lambda\right) -\frac{5}{2}\frac{k_B}{ \left| e \right|}\delta\lambda 
\label{eq:Ssc}
\end{eqnarray}
where the asymmetry parameter $\delta\lambda$ (that depends on $\mu$ and $T$) is given
by 
\begin{equation}
\delta\lambda=\frac{\lambda^c-\lambda^v}{\lambda^c+\lambda^v}
\end{equation}
Hence, a large Seebeck coefficient can be achieved by an interplay of the gap, $\Delta$,
the anisotropy or asymmetry $\delta\lambda$ in the transport function
-- stemming from either the densities of states  ($m^0_{c,v}$),
different bandwidth narrowings ($m^*_{c,v}$), scattering amplitudes
($\Gamma_{c,v}$) or  quasiparticle weights ($Z_{c,v}$).

\subsubsection{Upper limit for thermopower in a semiconductor}

The position of the chemical potential plays an important role in
maximizing the thermopower of a semiconductor. Let us take the example of
two equivalent bands. In this case, the term in brackets in
\eref{eq:Ssc} becomes
\begin{eqnarray}
\mu-\Delta/2\tanh(\beta\mu)
\label{ooTf}
\end{eqnarray}
This coefficient of the $1/T$ behavior is displayed in
\fref{figCmodel}(a) as a function of the chemical potential for a few 
different temperatures.
At high temperatures the optimal chemical potential,
which maximizes thermopower, is near the gap edges. At low temperature
$k_BT<\Delta/2$, the $1/T$ coefficient shows a local extrema.
The Seebeck coefficient vanishes at the point of particle-hole
symmetry ($\mu=0$ in the symmetric case considered here~
\footnote{We note that (even in the absence of impurities) the point
  of charge neutrality and the particle-hole symmetric point for the
  thermoelectric response are distinct from each other in the
  asymmetric case, in particular the latter is largely influenced by
  the Fermi velocities.}), and the optimal location of the
chemical potential at low temperature is in the direct vicinity
of that point, hence very close to the center of the gap.

\fref{figCmodel}(b) displays the value of the $1/T$ coefficient in
this extremum and the value at the gap edge as a function of
temperature. For $k_B T \gtrapprox 0.3 \Delta$ the maximum value of
the thermopower is achieved when the chemical potential is at the gap
edge, and for lower temperature, it is achieved close to the middle of
the gap, where the thermopower can reach the maximum value of
$S=\Delta/(2eT)$.

In an asymmetric case, the thermopower can be larger than this maximum
value, however, for a given charge gap $\Delta$, there is always an
upper bound for the Seebeck coefficient, namely
\begin{equation}
\left|S(T)e\right|\le \Delta/T + 5/2k_B.
\label{Smax}
\end{equation}
This is because the asymmetry is bounded to an absolute
value of one $\left|\delta\lambda\right|\le 1$.  This extremal value
corresponds to the fictitious system in which only one type of the two
carriers contributes to the thermoelectricity, e.g.\ the conduction electrons, and with the chemical
potential being, in that case, at the edge of the valence band.

Thus the correlation effects, such as small $Z$ in the conduction band
and large $Z$ in the valence band, can enhance the thermopower of a
semiconductor. However, this effect is limited by the form of
Eq.~(\ref{eq:Ssc}) allowing maximum $S$ bounded by
Eq.~(\ref{Smax}).  The possible merits of electron-hole-asymmetry for
the case of metals is discussed in Ref.~\onlinecite{haule_thermo}.

\subsubsection{Model semi-conductor in the presence of donor impurities}

\begin{figure*}[t!h]
  \begin{center}
   \hspace{-0.4cm}
\subfigure[\, effect of the impurity concentration $n_D$ on $\mu$, $S$, and $ZT$. ($\Delta$$=$$0.2$eV, $E_D$$=$$95$meV, $\Gamma$$=$$5\mu$eV)]
 			{\scalebox{1.}{\hspace{-0.2cm}\includegraphics[angle=-90,width=0.51\textwidth]{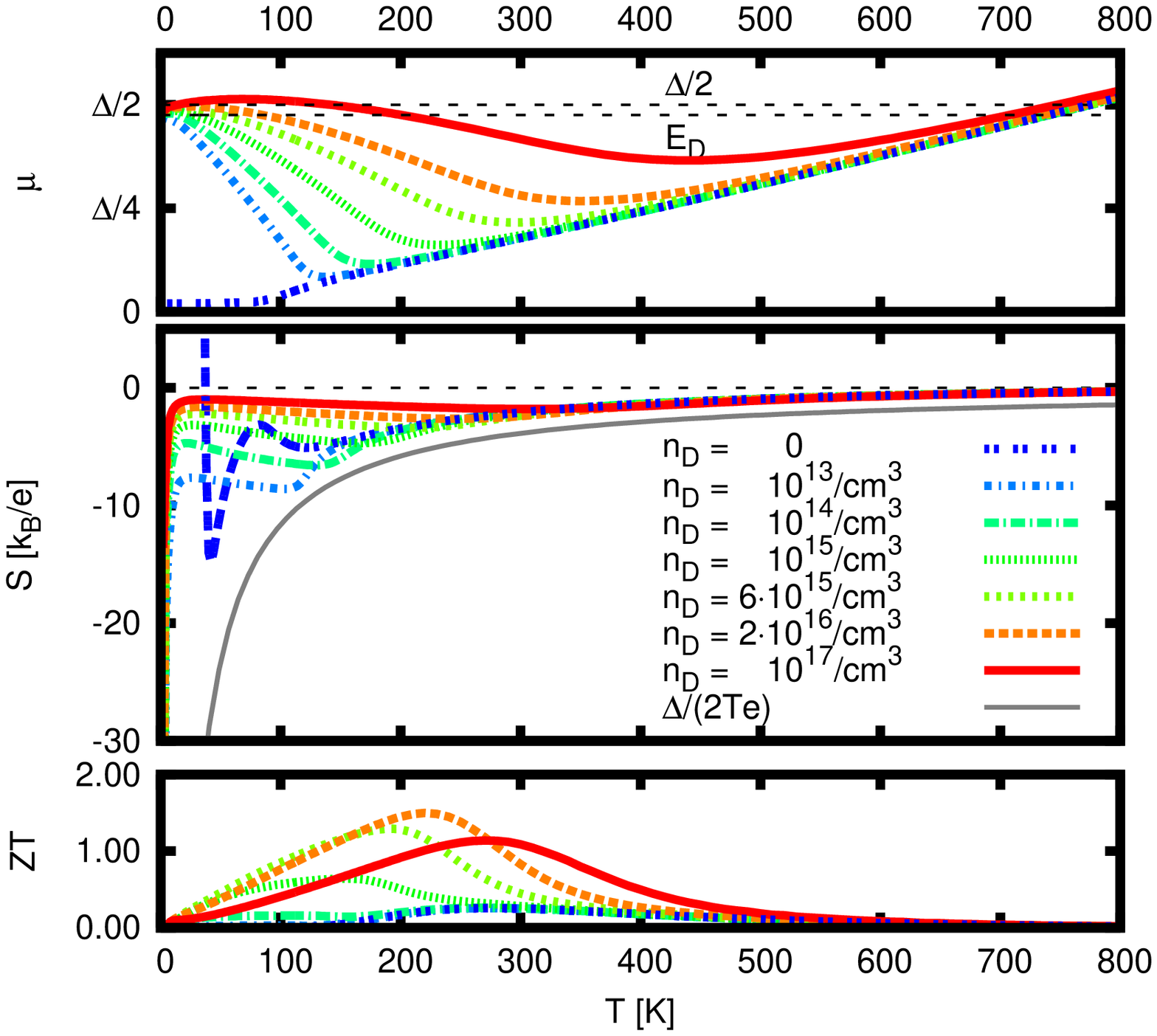}}} 
\hspace{-0.cm}
\subfigure[\, carrier density dependence of $S$, $\sigma$, and $\kappa$. ($T$$=$$220$K, $\Delta$$=$$0.2$eV, $\Gamma$$=$$5\mu$eV). Inset: Sketch
  of the quasiparticle dispersion. ]
			{\scalebox{1.}{\hspace{-0.2cm}\includegraphics[angle=-90,width=0.51\textwidth]{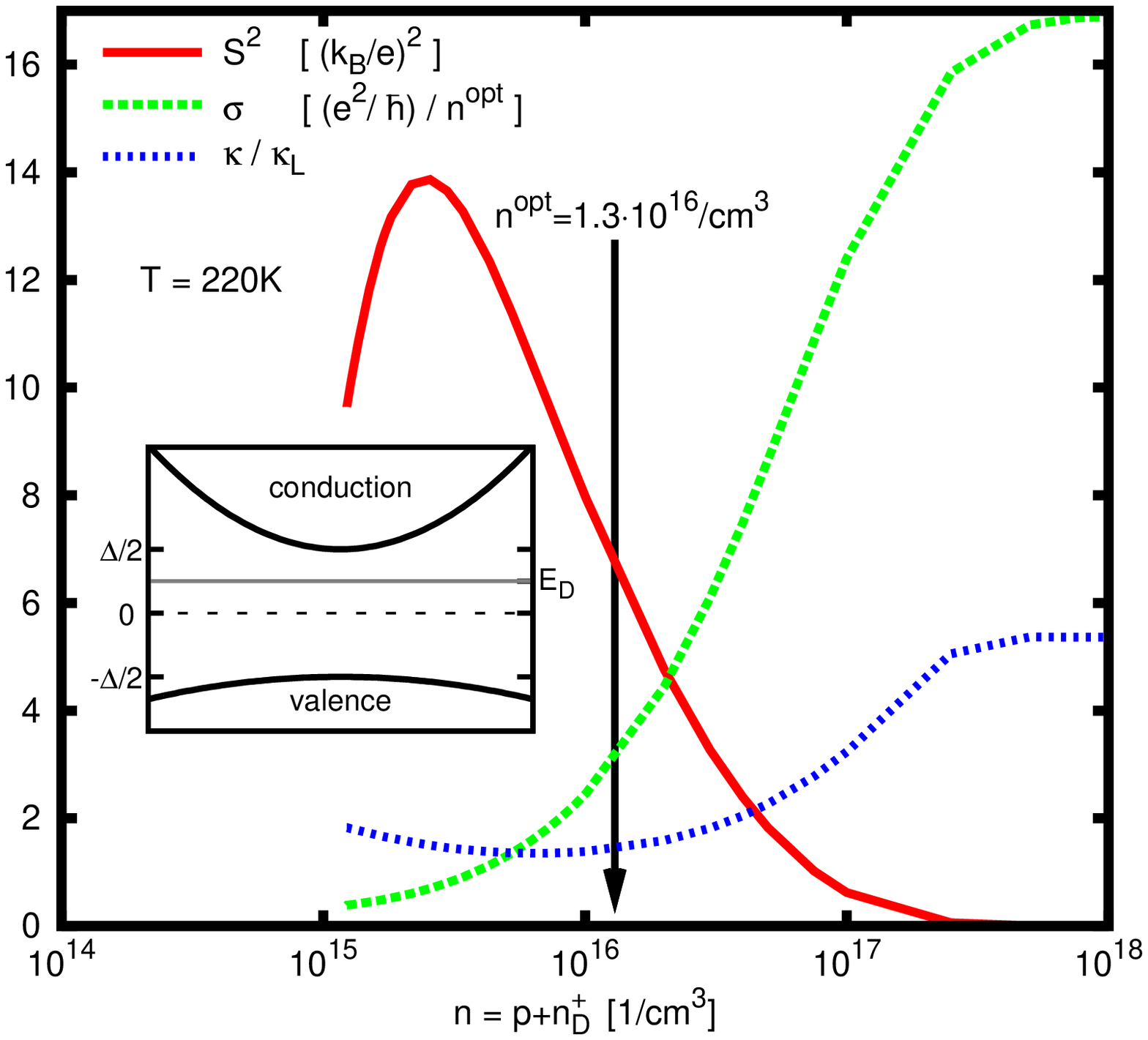}}} 

\hspace{-0.4cm}
\subfigure[\, effect of the scattering amplitude $\Gamma$ (top) and
  the position $E_D$ of the impurity (bottom). ($\Delta$$=$$0.2$eV, $n_D$$=$$2\cdot 10^{16}/$cm$^{-3}$). Inset: $S(T)$ for various scattering rates.]
			{\scalebox{1.}{\hspace{-0.2cm}\includegraphics[angle=-90,width=.51\textwidth]{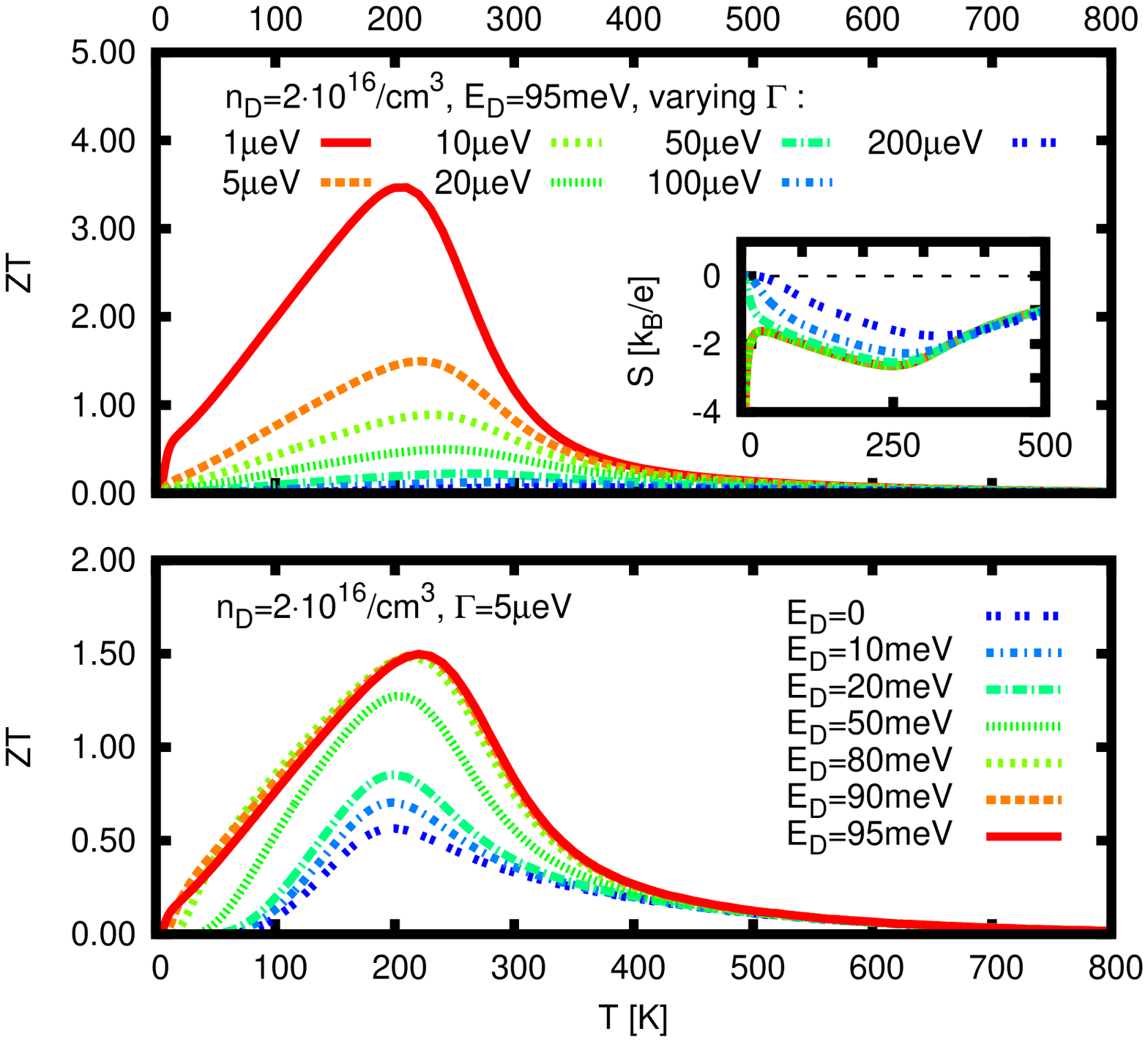}}} 
\hspace{-0.cm}
\subfigure[\, $ZT$ with and without lattice contributions (top), carrier concentration that maximizes $S$ or $ZT$ (bottom left), $ZT$ as a function of $n_D$ (bottom right). ($\Delta$$=$$0.2$eV, $\Gamma$$=$$5\mu$eV)]
			{\scalebox{1.}{\hspace{-0.2cm}\includegraphics[angle=-90,width=.51\textwidth]{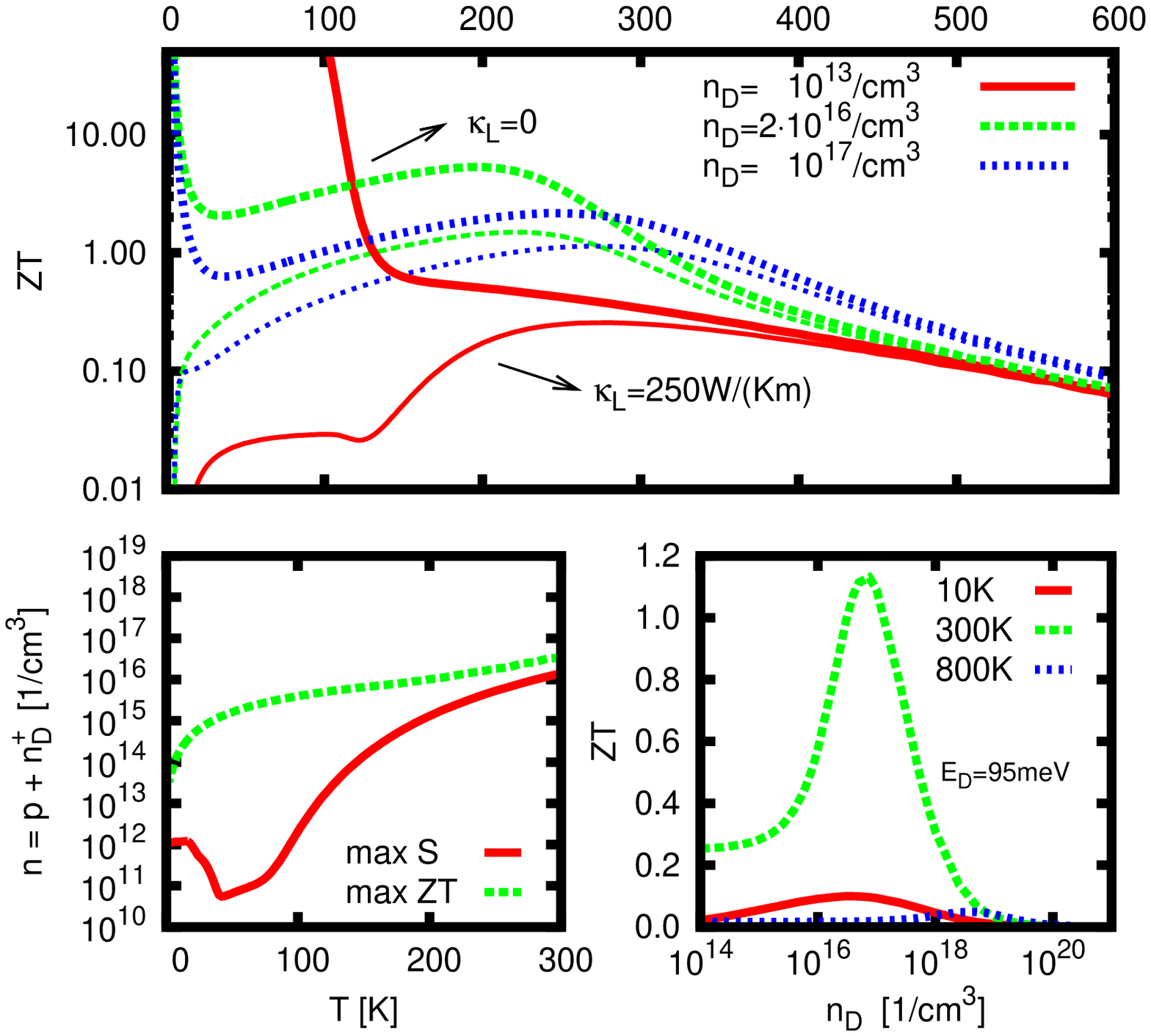}}} 
      \caption{
      Thermoelectric properties of the model semi-conductor with parameters inspired from \feas.
      (a) from top to bottom : 	chemical potential, Seebeck coefficient, and figure of merit $ZT$ as a function of temperature for different impurity concentrations. 
      (b) ingredients to the figure of merit $ZT=S^2\sigma T/\kappa$ at $T=220K$ as a function of carrier density. 
       The inset shows the general setup of asymmetric valence and conduction dispersions, with an impurity level $E_D$ in the gap $\Delta$. 
      (c) effect of the scattering amplitude $\Gamma$ (top) and the impurity level position $E_D$ (bottom) onto the figure of merit $ZT$, and the thermopower (inset).
      (d) top : comparison of the figure of merit from (a) with the purely electronic figure of merit ($\kappa_L=0$), bottom left : carrier density that maximizes $\left| S\right|$ and $ZT$ (for the chemical potential in the upper half of the gap), bottom right : $ZT$ as a function of the donor concentration for several temperatures, and a fixed impurity level $E_D=95$meV. We assume a unit-cell volume $V=80\AA^3$, and a thermal lattice conductivity (scaled with the quasi-particle weight $Z$, see text.) of  $\kappa_L/Z^2=250$W/(Km).}
      \label{model}
      \end{center}
\end{figure*}

With the goal of understanding the thermopower and the figure of merit
in a renormalized semiconductor in a very general setup, including the
presence of impurities, we now numerically study the model based on the
response functions \eref{XXX}. As before, we assume parabolic
dispersions \eref{eqxi}, with the band structure depicted in the inset
of \fref{model}(b)~: excitations of different effective masses are
separated in energy by a gap $\Delta$, and we allow for the presence
of donor impurities, situated at an energy $E_D$, as measured from the
middle of the gap.  We again assume transition matrix elements to be
given by the group velocity, \eref{groupvelo}.

We choose the parameters compatible with the band structure of \feas:
we consider a gap $\Delta=0.2$eV, and, unless stated otherwise, an
impurity level at $E_D=95$meV, as inferred from the low temperature
activation behavior of the resistivity\cite{APEX.2.091102}.  In our
current treatment, we assume that the impurity carriers have vanishing
Fermi velocities, and thus their only effect is to shift the chemical
potential.

To fix the particle--hole asymmetry, we note that at high
temperatures, the number of ionized impurities are irrelevant with
respect to the number of conduction and valence carriers, and the
chemical potential follows the intrinsic behavior. In the coherent
limit of the large gap semiconductor (see above) one finds that
\begin{equation}
\mu=3k_BT/4 \ln(\eta_v/\eta_c) \hbox{,$\quad$ with   } \eta=m^*/m_0\hbox{.} 
\label{muintrinsic}
\end{equation}
In this regime the resistivity shows an activation law with the
activation gap $\Delta/2$.  Of course, the situation in a real
material can be much more complicated (several types of impurities,
temperature dependence of the gap, etc.).
Using our {\it ab initio} data for \feas\ (that is presented below in
\sref{estruc}, and \sref{Sfeas2}), and assuming the approximate
validity of \eref{muintrinsic} for non-parabolic dispersions, one
finds the ratio of the valence and conduction effective mass
$\eta_v/\eta_c=2.5$ for \feas, which we will use for all the following
model calculations.

Further, we note that a uniform weight-factor $Z$ cancels in
the Seebeck coefficient, whereas in the figure of merit it can be seen as a scaling factor of the thermal lattice conductivity,
for which we assume  $\kappa_L/Z^2=250$W(Km)$^{-1}$.%
\footnote{This is a typical order of magnitude for our compounds of
  interest, see e.g.\cite{APEX.2.091102}. For simplicity we are
  neglecting the sizable temperature dependence
of the thermal lattice conductivity.}
Moreover, we use a unit-cell volume of $80\AA^3$.

Having thus fixed the size of the gap and the asymmetry, the principle
parameters to vary in this setup are the concentration of impurities
$n_D$ and the scattering rate $\Gamma$ (that we assume to be orbital
independent).  We will also study the dependence on the position of
the impurity band $E_D$ away from the value motivated by the
experimental resistivity.

\paragraph{impurity concentration.}
With these parameters, we display in \fref{model}(a) the temperature
dependence of the chemical potential, the Seebeck coefficient, and the
figure of merit for various impurity concentrations $n_D$, and for a
constant scattering rate $\Gamma=5\mu$eV.

In the intrinsic case, $n_D=0$, the chemical potential is indeed linear above a certain temperature
that is related to the scattering rate $\Gamma$. Below this regime, the chemical
potential is almost temperature independent. Since $\Gamma$ is small, the point of charge neutrality at zero temperature is
very close to the midgap point (where it is in the coherent case). The point of particle-hole compensation for the thermopower,
however, is higher in energy than the midgap point, as rationalized, above, for the symmetric semiconductor.
As a consequence, the chemical potential is actually moving through the symmetry point of the Seebeck coefficient,
and we expect it to change sign as a function of temperature. This is clearly seen in \fref{model}(a).

At finite dopings, and at low temperatures, the donor level plays the role of the valence
band, and the chemical potential (top panel) is between $E_D$ and the
conduction band.  In a coherent semiconductor the chemical potential
is $\mu\approx (\Delta/2 -\delta/2) -k_BT\ln(n/n_D^+)$, where
$\delta=\Delta/2-E_D$ is the impurity activation energy, seen in the
resistivity of this regime ($\rho\sim e^{-\beta\delta/2}$). At very
low temperatures, $\mu$ is pinned to $E_D$.  At intermediate
temperatures the chemical potential smoothly connects with the
intrinsic high temperature slope, as shown in \fref{model}(a).
For large enough concentration of impurities ($n_D>10^{16}/cm^3$), the
chemical potential can even go inside the conduction band at some
intermediate temperature, which can result in a shoulder, or even a
metallic slope in the resistivity (cf. the mentioned transport
measurements on \fesb, and \feas\cite{APEX.2.091102}).

For finite impurity concentration ($n_D>0$), the Seebeck coefficient
displayed in \fref{model}(a) may or may not be enhanced at a
given temperature, depending on whether or not the additional carriers
bring the chemical potential closer to its optimal value.
As explained above for the case of two particle hole symmetric bands,
at fixed temperature there exists a value of the chemical potential,
which maximizes the Seebeck coefficient.  In our asymmetric setup with
$\eta_v/\eta_c > 1$, the optimum chemical potential is located above
the mid gap point. In the limit of vanishing impurity concentration,
the midgap remains the point of charge neutrality at zero temperature,
while the optimal chemical potential is very near that point, hence
the Seebeck coefficient is a very strong function of temperature in
this limit, and can even change sign, as seen in the middle panel of
\fref{model}(a).
We note however, that the fundamental extremum, established by the
size of the gap, is always respected as is evident in
\fref{model}(a).

The efficiency of the thermoelectrical material is determined by its
figure of merit $ZT$, which we also plot in the lower panel of
\fref{model}(a).  $ZT$ can be greatly enhanced by the presence of
impurities,  and its
maximum is not necessarily in close vicinity of the thermopower
maximum.
 Indeed the largest $ZT$ for
the current parameters is achieved at about $T=220$K for a
concentration $n_D=2\cdot 10^{16}$/cm$^3$. 
The Seebeck coefficient at the
this point is actually smaller than in the intrinsic limit.
The position of the impurity with respect to the gap edge has a large
effect on the optimal impurity density. While for our specific choice
of parameters, $\Delta/2 - E_D = 5$meV, the optimal density is $n_D =
2\cdot 10^{16}$/cm$^3$, we notice that for larger gaps and/or larger separation of
the impurity level from the gap edge, the optimal impurity density can
reach a value as large as $\sim 10^{20}$/cm$^3$.\cite{mahan:42}
 To
elucidate the origin of an optimal density further, we show in \fref{model}(b) the dependence of
the quantities entering the expression of $ZT$ as a function of the
particle density ($n=p+n_D^+$) at fixed $T=220K$, the temperature which
maximizes $ZT$ in \fref{model}(a). We use here the total particle
density, because in this case the description becomes independent of
the bare impurity concentration and the level position, and in
particular the additional carriers can find their origin from multiple
impurity sources.

For the given gap, and $T=220K$ no smaller densities than
$10^{15}$/cm$^3$ can be accessed.  The optimum $n_D$ found in
\fref{model}(a) translates into $n=1.3\cdot 10^{16}$/cm$^3$, mainly as
a trade-off between decreasing $S^2$ (less entropy per carrier) and
increasing $\sigma$ (larger conductivity with more carriers). In
particular, we note that $S$ achieves its maximum for smaller
concentration of carriers then the figure of merit.

The thermal conductivity in this range varies very slowly with
concentration, which is not surprising since we fixed the lattice
contribution to thermal conductivity to a fixed value of
$\kappa_L/Z^2=250$W(Km)$^{-1}$.

Having chosen parameters to represent \feas, we note that the
experimentally measured carrier concentration in this compound, as inferred from Hall measurements,
is $5\cdot 10^{17}$/cm$^3$ in the range of 60-170K,~\cite{Fan1972136}
which is higher than the density that optimizes $ZT$ in our model.
Thus, it seems conceivable that by a deliberate change in the impurity concentration or position,
an increase in the figure of merit of the specimen can be achieved.

\paragraph{lifetime effects.}
Next we pick the impurity concentration which maximized $ZT$, and we
investigate the role of the scattering rate for the figure of merit
and the Seebeck coefficient in \fref{model}(c).
The life-time has two effects: through the change in chemical potential, 
and directly through the dependence of the response functions
\eref{XXX} on scattering rate $\Gamma$.  It is this latter effect that
causes the Seebeck coefficient to vanish at low temperatures for a
sufficiently large scattering amplitude, as can be seen in the inset
of the top panel of \fref{model}(c). The increase in scattering rate
reduces the absolute value of both the Seebeck coefficient and the
figure of merit, hence long lifetimes are preferred in thermoelectric
materials.

The upper limit of the Seebeck coefficient has been discussed
above. As a function of the scattering rate $\Gamma$, figure of merit
$ZT$ is limited as well. If lifetimes are long, the dependence
of the response functions thereof is linear, and thus cancels in
the dimensionless ratio $ZT$ if there are no lattice contributions to
the thermal conductivity.  Therefore, with decreasing $\Gamma$, the
figure of merit converges towards the purely electronic limit in which
$\kappa_L=0$.

\paragraph{position of the impurity level.}

In \fref{model}(c) we show the dependence of the figure of merit
on the position of the impurity level $E_D$. We fix the impurity
concentration to $n_D=2\cdot 10^{16}$/cm$^3$ and the scattering rate to
$\Gamma=5\mu$eV.  It is clear from \fref{model}(c) that a maximum
$ZT$ is achieved when the donor level is very close to the conduction
band, which is located at $\Delta/2=100$meV.

This can again be understood from the optimal number of carriers~:
Indeed, in order to reach the ideal electron density of $1.3\cdot
10^{16}$/cm$^3$, the chemical potential must be rather close to the
conduction band.  Since at $T=200$K and a gap of $\Delta=0.2$eV, the
compensating holes cannot come from the valence electrons, they have
to be supplied from the impurity band, hence, for an impurity
concentration $n_D=2\cdot 10^{16}$/cm$^3$ that is larger than the
needed number (if completely ionized), also $E_D$ must be very close
to the conduction band.

Next we study the sensitivity of the figure of merit to the lattice
thermal conductivity. With thick line in \fref{model}(d) we show the
$ZT$ in the absence of lattice thermal conductivity $\kappa_L$ and
with a thin line is shown $ZT$ for a constant value of
$\kappa_L/Z^2=250$W/(Km). The figure of merit is clearly enhanced when
the lattice conductivity is reduced, hence the desire for a ``phonon
glass'' (see e.g.\ the review \cite{Snyder2008}), i.e.\ a solid which
has a low phonon mean free path such as to prevent substantial heat
conduction by lattice vibration modes.  The effect of lattice thermal
conductivity is most enhanced at low temperature and for small
impurity concentration.

In \fref{model}(d) lower panel we plot the optimal carrier
concentration, which maximizes $S$ or $ZT$, as a function of
temperature. The $ZT$ curve is monotonically increasing function of
temperature, hence for best performance at higher temperature we need
more impurity carriers. To maximize $ZT$ we need larger impurity
concentration than we need to maximize $S$.  Finally, we also display
the value of $ZT$ as a function of impurity concentration for a few
representative temperatures. For our setup of parameters, the figure of
merit is very sharply peaked at room temperature around a 
carrier concentration ($n_D\approx 10^{17}$/cm$^3$).


In conclusion, these model considerations give guidance as to where to
look for promising thermoelectric materials.  In particular we showed
that -- unless vertex corrections or strongly frequency dependent lifetimes are of pivotal importance --
electronic correlation effects are not in the position to enhance the
thermopower of a gapped system, they can only shift the asymmetry of
the contributions for electrons and holes. Indeed we found the
thermopower (of purely electronic origin) to have an upper bound that
is given by the size of the gap.


\section{Electronic Structure of \fesb\ and \feas}
\label{estruc}

\begin{figure*}[t!h]
  \begin{center}
    \mbox{
\hspace{-1.5cm}
\subfigure[\, \fesb\ : GGA]{\scalebox{0.35}{\includegraphics[angle=0,width=1.\textwidth]{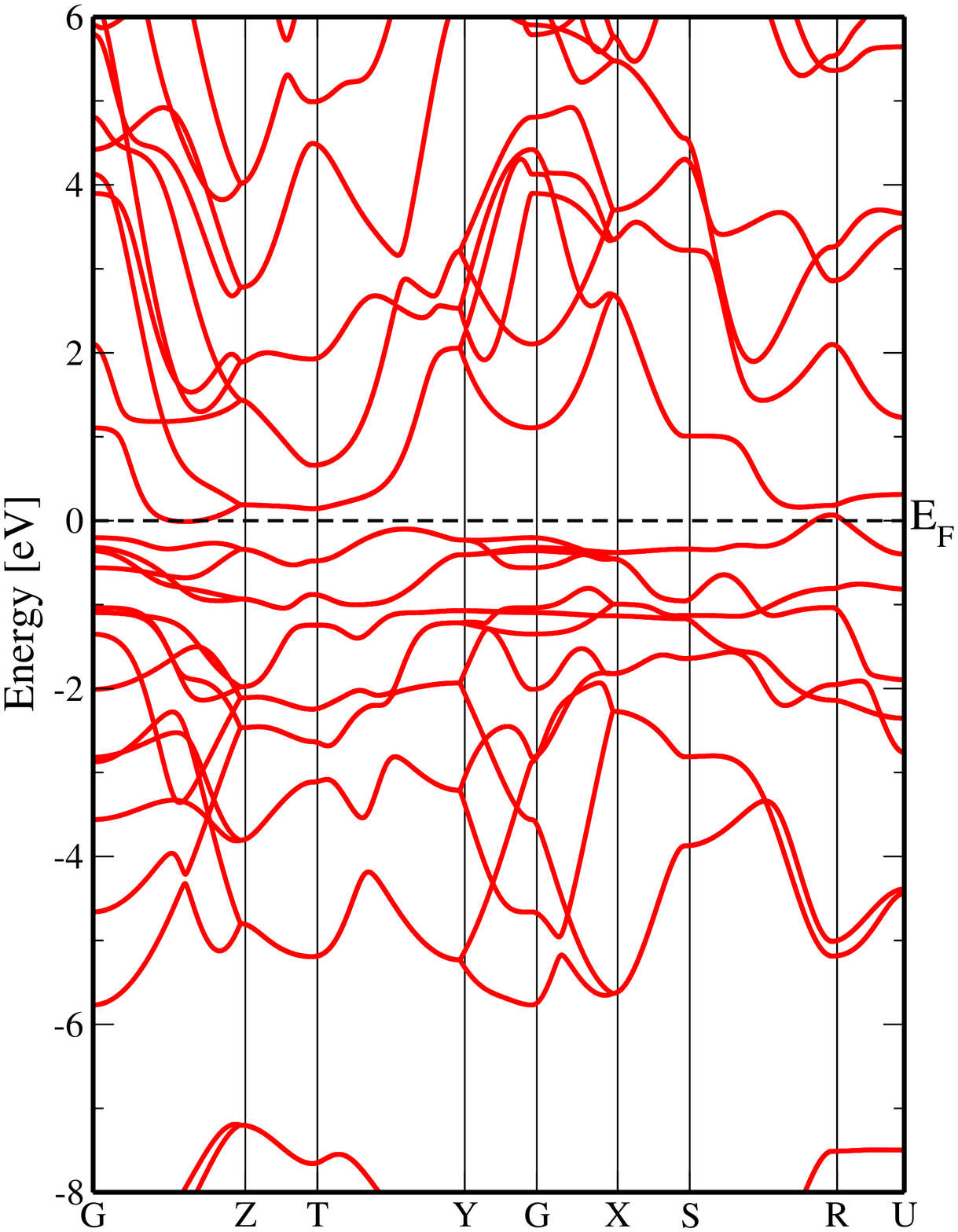}}} 
\hspace{-.25cm}
\subfigure[\, \fesb\ : Hybrid functional]{\scalebox{0.35}{\includegraphics[angle=0,width=1.\textwidth]{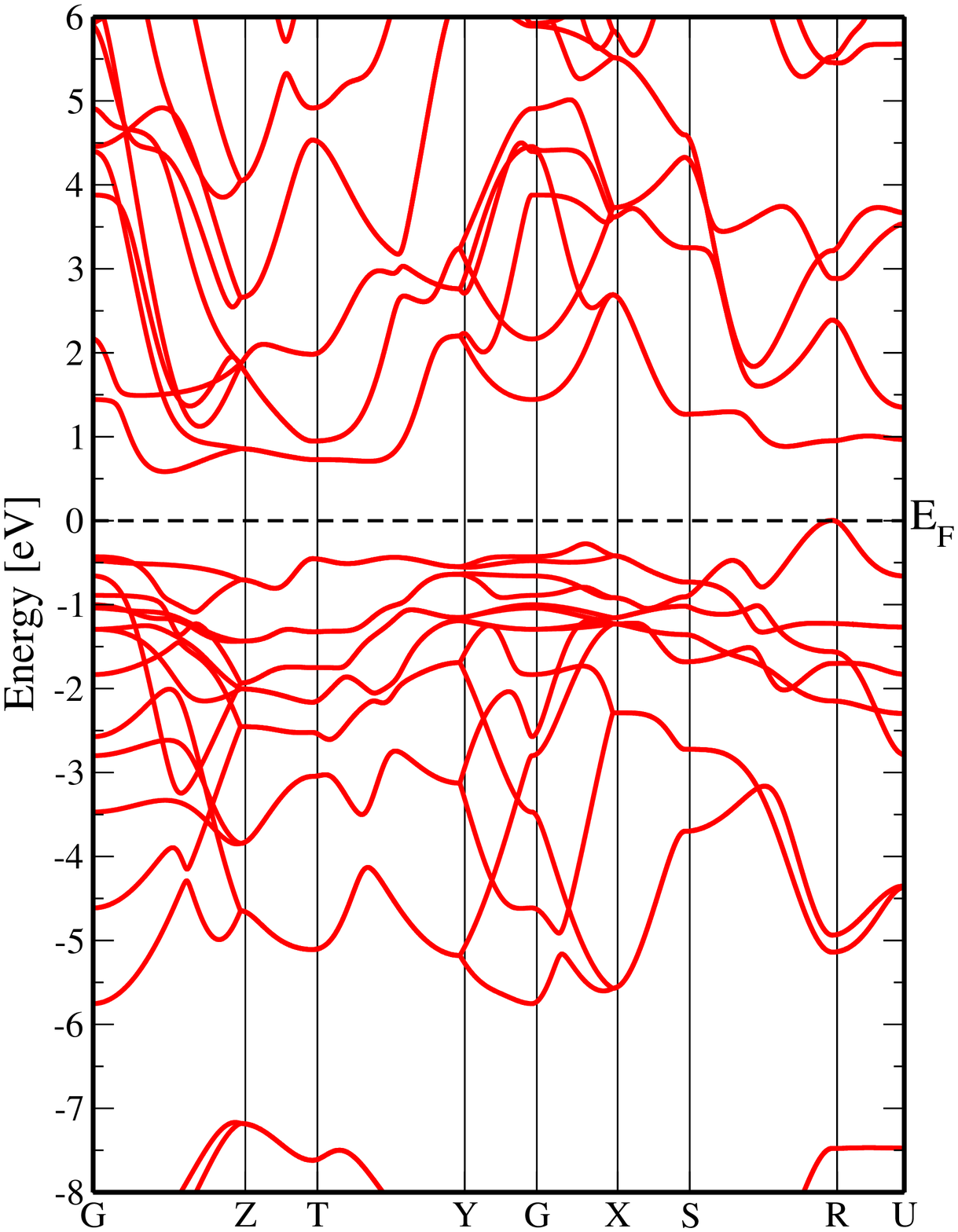}}} 
\hspace{-.25cm}
\subfigure[\, \feas\ : GGA]{\scalebox{0.35}{\includegraphics[angle=0,width=1.\textwidth]{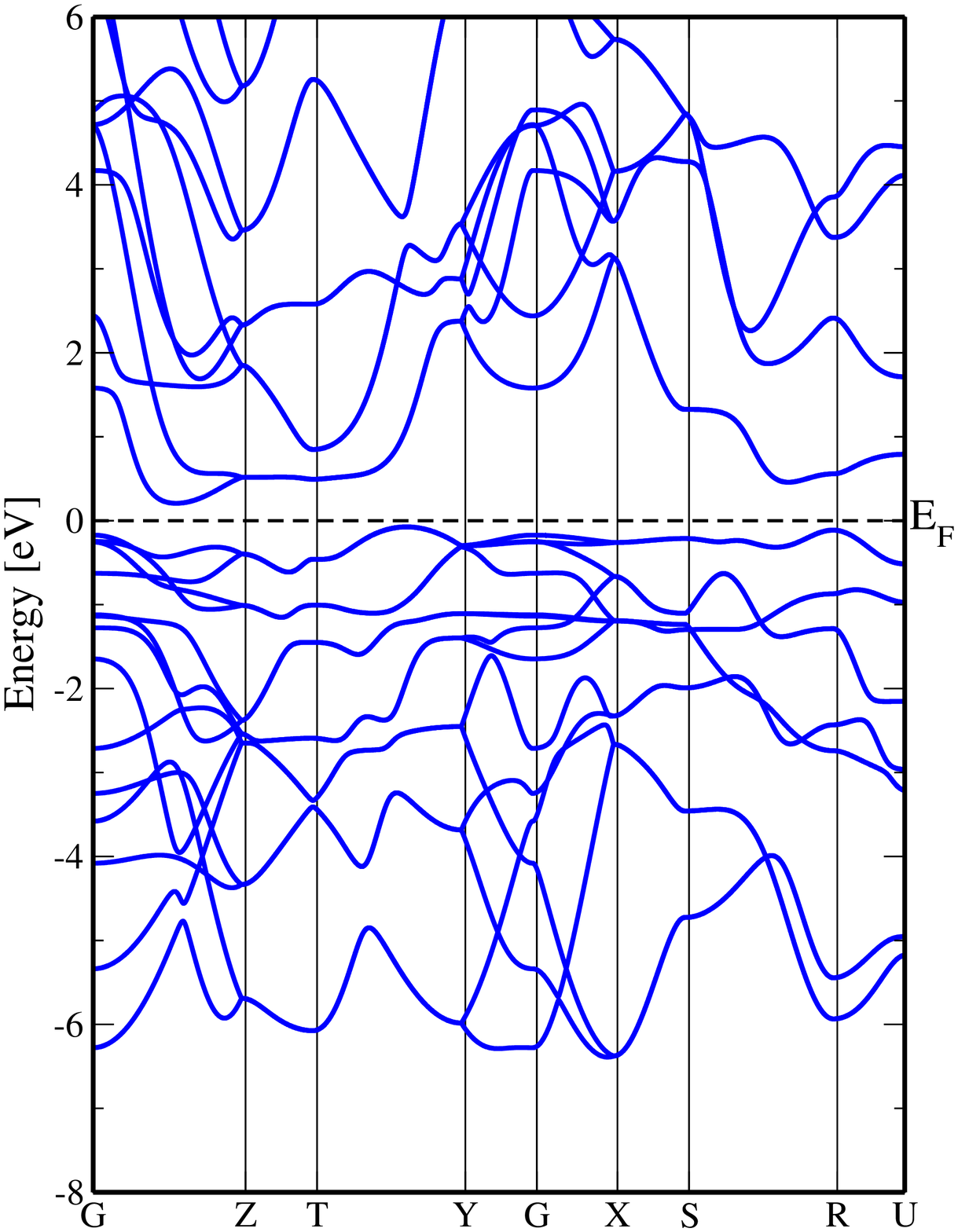}}} 
      }
      \caption{Band-structures of FeSb$_2$ (left : GGA, middle : Hybrid functional) and FeAs$_2$ (right : GGA). }
      \label{fig1}
      \end{center}
\end{figure*}

\fesb\ and \feas\ both crystallize in the (regular) marcasite structure, have the orthorhombic space group Pnnm, and there are two formula units per unit cell (see Ref.~\onlinecite{Goodenough1972144}).
The iron ions are surrounded by distorted pnictogen octahedra, that
share corners along the c-axis (see e.g. Ref.~\onlinecite{PhysRevB.72.045103}).
In the ligand-field picture, the compounds have a Fe3d$^4$ configuration and 
the 3d-orbitals are split into $e_g$ and lower lying \t2g orbitals. The inequivalence of Fe--pnictogen distances causes the \t2g\ to split further
into two degenerate lower and one higher orbital.
In this picture, the compounds are in an insulating low spin state with the two degenerate \t2g\ orbitals filled.\cite{Goodenough1972144}
From this perspective,
metalization of \fesb\ is driven by a temperature induced population of the third \t2g\ orbital.\cite{Goodenough1972144,PhysRevB.72.045103}
Previous band-structure calculations, however, suggest a more covalent picture~\cite{Madsen_fesb2}, in the sense that stabilization occurs for d-orbitals that point
towards the ligands, i.e.\ in particular lowering the $e_g$ orbitals. 
In \fesb\ this happens to the extend that LDA calculation actually yield a metallic ground-state~\cite{luko_fesb2,bentien:205105,Madsen_fesb2}$^,$\footnote{ In Ref.~\onlinecite{luko_fesb2} a gap of 0.3eV was extracted from the density of states. Yet, the displayed bandstructure is metallic. 
}.

From the perspective of electronic structure methods, at zero temperature, the challenge is hence to obtain an insulting ground state for \fesb.
We therefore compare  how three different approaches, GGA, Hybrid functionals and the GW approximation perform in this problem.
In preparation for future work which should include correlations beyond GW, and to clarify  what would be the starting Hamiltonian to describe these materials,  
we obtain transfer matrix elements and estimate the values of the interaction using the constrained random phase approximation (cRPA) method.

\paragraph{band-structures.}
Our results for the  band-structure of \fesb\ given by the GGA~\cite{PhysRevLett.77.3865} 
of DFT as implemented
in the Wien2k package\cite{wien2k} are displayed in  \fref{fig1}a.
We used the atomic positions at room-temperature (a=5.83\AA, b=6.54\AA, c=3.20\AA)~\cite{fesb2_struct}$^,$\footnote{
Results (not shown) for low temperature positions~\cite{PhysRevB.72.045103} do not qualitatively differ as far as the following discussion is concerned.
}%
.
Congruent with previous works~\cite{luko_fesb2,bentien:205105,Madsen_fesb2}, the GGA ground-state is metallic with small electron pockets half-way between the $\Gamma$ and $Z$
symmetry points, and corresponding hole-pockets at all corners, $R$, of the orthorhombic Brillouin zone. 
Crucial to the understanding of the gap mechanism within more sophisticated techniques (see the GW discussion below) is to note the 
different orbital characters of the pockets.\cite{luko_fesb2,Madsen_fesb2}
To quantify this, we transform the local coordinate system of the d-orbitals into a basis, in which the local projection of the d-block of the (GGA) Hamiltonian is as diagonal as possible.
In this coordinate system, the x and z axes point (almost) towards the antimonide atoms,\footnote{%
The new unit vectors are $x=(-0.627 ,  0.326,  0.707)$, $y=( 0.627 , -0.326,  0.707)$, $z=( 0.462,  0.887,  0.        )$.}
and the $e_g$ orbitals exhibit the expected bonding/anti-bonding splitting. 
In this basis, the electron pocket is mainly of $d_{xy}$ character, and the hole pocket is formed by the now degenerate $d_{xz}$ and $d_{yz}$ (\t2g) orbitals.
The respective occupations (within the muffin spheres) are shown in \tref{tabn}.

\begin{table}%
\begin{tabular}{c|ccccc}
\fesb & $\phantom{-}d_{z^2\phantom{y^2}}$ & $d_{x^2-y^2}$ & $\phantom{-}d_{xz\phantom{y^2}}$ & $\phantom{-}d_{yz\phantom{y^2}}$ &  $\phantom{-}d_{xy\phantom{y^2}}$ \\
\hline
n & 1.27& 1.42&1.42 &1.42 & 1.06\\
\end{tabular}
\caption{Occupations of the $d$-orbitals (in the muffin spheres) within the transformed local coordinate system (see text for details). The \t2g\ orbitals $d_{xz}$ and $d_{yz}$ are degenerate and mainly account for the hole pocket. The electron pocket is of $d_{xy}$ character.}
\label{tabn}
\end{table}

The GGA band-structure of \feas\ (we use a=5.3\AA, b=5.98\AA, c=2.88\AA \cite{Fan1972136}) 
is shown in
\fref{fig1}c. With respect to \fesb, the chemical pressure of the larger As atoms is almost isotropic, and the c/a ratio remains virtually constant
(as a function of external pressure, the ratio slightly decreases\cite{PhysRevB.72.045103}). 
In consequence, and as is apparent from the graph, the bands of \feas\ are much akin to those of \fesb, and could have roughly been obtained from a rigid band-shift%
\footnote{We note that the former hole pockets on the Brillouin zone corners are slightly surpassed as the uppermost valence band by momentum regions on the way from the midpoint of the xz-face ($Y$-point), up z-wards towards the edge point $T$.}%
.
While within GGA the gap of \fesb\ was underestimated (no gap at all), the value $0.28$eV for \feas\ is just slightly too large with respect to the experimental $0.2-0.22$~eV~\cite{Fan1972136,APEX.2.091102}. 
On a qualitative level, one could thus say that a DFT calculation seems to work rather well for \feas.
Below we will explain why we believe this to be a mere coincidence.

\begin{figure*}[t!h]
  \begin{center}
\subfigure[\, \fesb\ ]{\includegraphics[angle=-90,width=.45\textwidth]{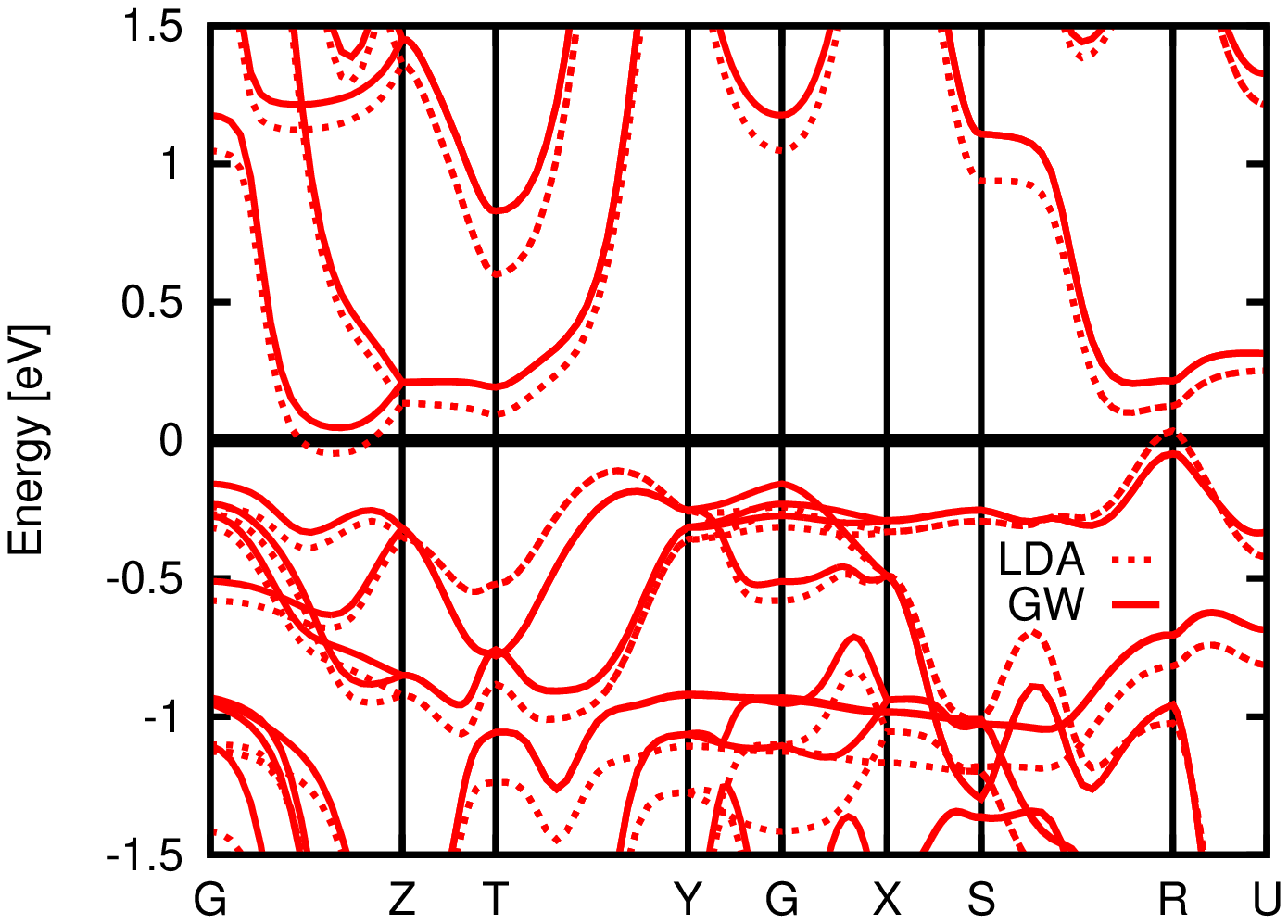}}
\subfigure[\, \feas\ ]{\includegraphics[angle=-90,width=.45\textwidth]{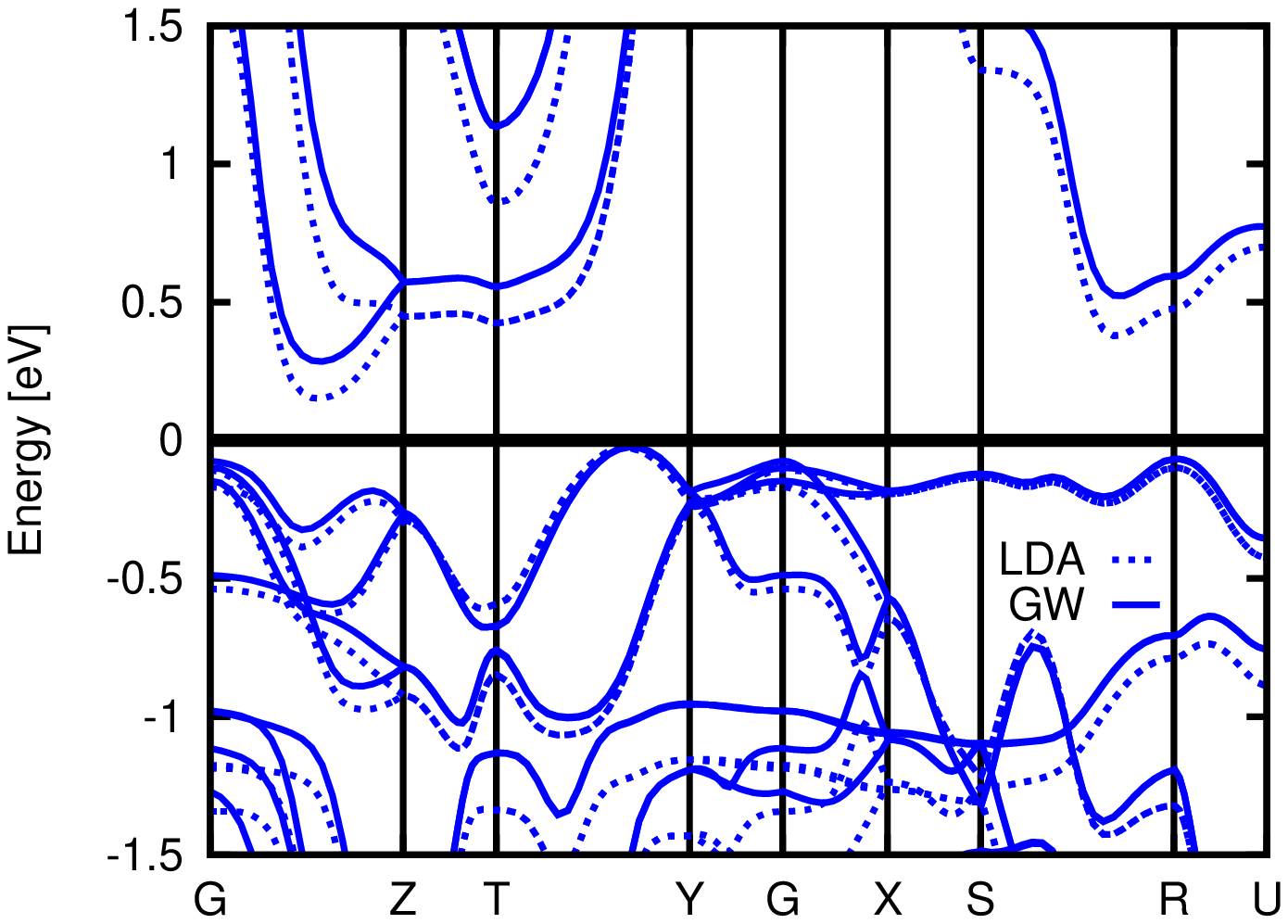}} 
      \caption{band-structure in the GW approximation (full lines), in comparison with LDA (dashed).}
      \label{fig1b}
      \end{center}
\end{figure*}

\paragraph{Maximally localized Wannier function and cRPA.}
From a conceptual point of view, we find it insight-full to note, and compare, the hierarchy of transfer matrix elements and the magnitude of local Hubbard interactions.
Starting from a full-potential (FP) LMTO (LDA) computation~\cite{fplmto}, we construct
maximally localized Wannier functions for the subsystem consisting of the Fe-3d and Sb-5p orbitals, as described e.g.\ in Ref.~\onlinecite{PhysRevB.56.12847,PhysRevB.65.035109,miyake:085122}, 
and find that the (largest) nearest neighbor hopping amplitudes in
\fesb\  are 
 $t_{dd}\sim 0.2-0.3$eV,  $t_{pd}\sim 0.95$eV, and
 $t_{pp}\sim 0.7$eV. The significance of both the Sb-5p dispersion and the large hybridization was heralded already in the band-structure, \fref{fig1}a and Refs.~\onlinecite{luko_fesb2,Madsen_fesb2},  as well as in
 the strongly mixed orbitals characters,~\cite{luko_fesb2,Madsen_fesb2} see also the recent work Ref.~\onlinecite{wien2wannier}.
The two transfers $t_{pd}$, $t_{pp}$ being of comparable magnitude, the system thus lies between the canonical Anderson model ($t_{pp}\gg t_{pd}$) and cuprate like compounds ($t_{pp}\ll t_{pd}$).
The centers of gravity of the $d$ and $p$ bands are separated by $\Delta_{ct}=4.2$~eV. 
In \feas\ the $pd$-transfers are larger, as expected from the chemical pressure and the covalent/hybridization-like character of the gap. We find
$t_{dd}\sim 0.25-0.3$eV, $t_{pd}\sim 1.1$eV, and $t_{pp}\sim 0.8$eV, and $\Delta_{ct}=5.0$~eV.

To investigate the strength of local interactions, we compute the Hubbard $U$ using the constrained RPA~\cite{PhysRevB.70.195104,miyake:085122} technique.
Since in \fesb\ and \feas\ the eigenvalues of the $pd$-subspace are entangled with higher energy bands, we employ the scheme presented in Ref.~\onlinecite{miyake:155134}.

In relation to the rather small energetic shift needed to deplete the (GGA) pockets in \fesb, we find, see \tref{U},  that 
the orbital dependence of the Hubbard $U$ within the $d$--orbitals is of notable 5\% -- an effect hitherto mostly neglected in methods for correlated materials that start from a parametrized  Hubbard-like Hamiltonian. 
Before, we alluded to the different orbital characters of the pockets.
Here, we indeed find for the $d_{xz}$ and $d_{yz}$ orbital that mainly account for the hole pocket a value of $U=8.6$~eV, while for the 
orbital corresponding to the character of the electron pocket the interaction is slightly larger with $U_{d_{xy}}=8.8$~eV. 
This differentiation in principle favors a charge transfer towards a gap opening.
Further matrix elements of the interaction are $U_{pd}\sim2.5$eV, and $U_{pp}\sim4$eV.
The hierarchy of the interaction strength within the d-shell is already seen in the bare, i.e.\ unscreened, Coulomb interaction
and is thus linked with the construction of the 
localized orbitals. 
On general grounds, 
larger matrix elements of the unscreened interaction are indeed expected for orbitals that have stronger hybridizations with other orbitals, thus are more spatially 
delocalized~\cite{jmt_wannier}.
In analogy with the pressure dependence of Coulombic interactions in a localized basis~\cite{jmt_wannier,jmt_mno}, the bare interaction of \feas\ is larger than that of \fesb,
in the cubic reference frame we find e.g.\ $V_{d{z^2}}=22.7$eV and $V_{d{x^2-y^2}}=22.6$eV (the corresponding values for \fesb\ are $V_{d{z^2}}=22.0$eV, $V_{d{x^2-y^2}}=21.5$eV). Moreover, with respect to the antimonide, the Kohn-Sham eigenvalues of the arsenide move towards higher energies, therewith reducing screening strengths, and causing significantly larger values also for the Hubbard $U$~: $U_{d{z^2}}=11.0$~eV, $U_{d{x^2-y^2}}=10.7$~eV. 

\begin{table}%
\begin{tabular}{l|ccccc}
\fesb & $\phantom{-}d_{z^2\phantom{y^2}}$ & d$_{x^2-y^2}$ & $\phantom{-}d_{xz\phantom{y^2}}$ & $\phantom{-}d_{yz\phantom{y^2}}$ &  $\phantom{-}d_{xy\phantom{y^2}}$ \\
\hline
$d_{z^2\phantom{y^2}}$  & 8.5 &  7.0 &  7.3 &  7.3 &  7.0 \\
$d_{x^2-y^2}$           & 7.0 &  8.8 &  7.2 &  7.2 &  7.5 \\
$d_{xz\phantom{y^2}}$   & 7.3 &  7.2 &  8.6 &  7.0 &  7.3\\
$d_{yz\phantom{y^2}}$   & 7.3 &  7.2 &  7.0 &  8.6 &  7.3\\
$d_{xy\phantom{y^2}}$   & 7.0 &  7.5 &  7.3 &  7.3 &  8.8\\
\end{tabular}
\caption{constraint RPA values for the Hubbard $U$ (in eV) of \fesb\ for the Fe3d-orbitals in the pd-setup of maximally localized Wannier functions in the local coordinate system.}
\label{U}
\end{table}

\paragraph{hybrid functional approach.}
Previous attempts to produce an insulating band-structure for \fesb\ were made within the
LDA+U scheme~\cite{PhysRevB.44.943}, where the paramagnetic state (LDA) was found to be stable below a critical U=2.6eV with respect to a ferromagnetically ordered phase (LDA+U)~\cite{luko_fesb2}. 
Here, we use a hybrid functional approach (HYB)~\cite{becke:1372}, and
\fref{fig1}b displays the resulting band-structure, where the B3PW91 functional was used for the $d$-orbitals of the iron atoms\footnote{In Wien2k, the hybrid functional approach
is implemented only within the atomic spheres. Therewith one can choose to apply corrections to specific atomic characters.}%
.
We also note that, while given the freedom, the system does not develop any magnetic moment within this setup (in LDA+U it necessarily does). 
The band-structure features an indirect gap of about 0.6eV, i.e.\ it is by far larger than in experiment.
This points towards, both, a serious underestimation of static correlations within the previously used GGA, and the 
lacking of dynamical effects in the hybrid functional approach that will work to reduce the size of the gap.

\paragraph{GW approximation.}
To investigate the dynamical effects of electronic correlations in our compounds, we applied Hedin's (non-selfconsistent) GW approximation~\cite{hedin}, which has proven to be quite successful for semi-conductors~\cite{ferdi_gw,RevModPhys.74.601}, in its FP-LMTO realization~\cite{fpgw} to both, \fesb\ and \feas.  
In \fref{fig1b}a,b we display, besides the FP-LMTO (LDA) Kohn-Sham energies $\epsilon_{KS}$, the band-structure obtained by taking into account (perturbatively) the energy shifts as provided by the GW self-energy 
\begin{eqnarray}
\epsilon_{GW}\approx Z\biggl[\epsilon_{KS}+Re\Sigma(\epsilon_{KS})\biggr]
\label{EGW}
\end{eqnarray}
with $Z^{-1}=1-\partial_\omega\left.Re\Sigma\right|_{\omega=\epsilon_{KS}}$.
In the case of \fesb\ this indeed opens a charge gap in agreement with experiment.
We note that, consistent with the above hybrid functional calculation, as well as with the discussed orbital-dependent interaction strength, 
the static part of the GW self-energy, i.e.\ setting $Z=1$ in \eref{EGW}, yields a too large gap of $\sim 0.2$~eV. 
Thus, it is the dynamics of the self-energy, therewith a true correlation effect, that scales down the gap size with respect to Hartree-like approaches --
a situation quite akin to that of correlated band insulators~\cite{kunes:033109,sentef:155116}.
Indeed, the real part of the diagonal matrix-elements of the self-energy are linear in frequency over an extended energy range of up to 10~eV.
While the derivative of these elements are basically orbitally independent within the Fe-3d, and Sb-5p orbital subsets, respectively, the different hybridizations and also the different
off-diagonal elements yield for the antimonide, in the Kohn-Sham basis, a minimal value (eigenvalue of the self-energy derivative matrix) of $Z\approx0.52$ for "bands" near the Fermi level, 
and $Z\approx 0.6-0.7$ for higher lying "bands"\footnote{We stress that the parameter $Z$ occurs here only formally as parameter in the frequency expansion of the self-energy, and should not be confounded with the quasi-particle weight in a Fermi liquid.}%
.
Concomitant with the linear slope of the real-part, the imaginary part of the self-energy is basically quadratic, but notably asymmetric with respect to the Fermi level, see \tref{gamma}.

Despite the larger values of the Hubbard $U$, the correlation dynamics is less pronounced in \feas, and values of $Z$ reach a minimum of $Z\approx 0.6$ for excitations closest to the Fermi level. Also, lifetime effects are both smaller in magnitude, and less asymmetric, see \tref{gamma}.
While for \fesb\ the GW approach has the correct trend with respect to experiments, the gap of \feas\ also slightly increases from its LDA value, thus departing a bit further from the experimental value of $\sim 0.2-0.22$~eV. 
We note that while the gap size within LDA/GGA and GW are comparable, the physics is not~: 
In the GW, the size of the gap is a result of an almost compensation between static (exchange-like) contributions (see hybrids) and the dynamical correlations. 
 The induced bandwidth narrowing within the GW distinguishes its excitations from the KS spectrum.

\begin{table}%
\begin{tabular}{l|c|c}
$\Gamma$ [eV$^{-1}$]&\fesb &\feas\\
\hline
$\omega<0$ &0.15 & 0.08 \\
$\omega>0$ &0.02-0.05 & 0.02-0.03\\
\end{tabular}
\caption{comparison of the asymmetry of the scattering amplitude within the GW approximation. Extraction by fitting the average d-orbital self-energy (in the Kohn-Sham basis) by $\Im\Sigma(\left|\omega\right|<5~\hbox{eV}) = -\Gamma\omega^2$.}
\label{gamma}
\end{table}

\section{Realistic Seebeck coefficients for \feas, and \fesb}

In the light of the above considerations for the thermopower of semiconductors, the principle puzzle now is why \fesb,
while having a gap that is about 7 times smaller than that of \feas, has a Seebeck coefficient that is (up to) 5 times larger.

For the calculation of the realistic Seebeck coefficient given in \eref{eqS}, we employ the Fermi velocity matrix elements of the optics implementation~\cite{AmbroschDraxl20061} of Wien2k,
and compute the correlation functions according to \eref{A01}.

\subsection{\feas}
\label{Sfeas2}
The band-structure underlying the theoretical Seebeck coefficient of \feas\ is the GGA result shown in \fref{fig1}c.
Since the size of the gap is important for the magnitude of the thermopower, we scale down its size from its GGA value of 0.28~eV
to the experimental 0.2~eV. 
The calculation of the transport coefficients uses a small, frequency and momentum independent scattering rate/self-energy ($\Gamma\sim 20$meV), which therewith practically cancels out in the Seebeck coefficient (see the discussion above),\footnote{%
There is however an apparent influence of the line broadening on the determination of the chemical potential. Lacking for finite k-mesh sampling (without the tetrahedron method) the numerical precision to avoid spurious in gap spectral weight, 
we extract from the realistic data the linear high temperature evolution, and, via \eref{muintrinsic}, use the realistic asymmetry in the semi-conductor model, \eref{muintrinsic}, (along with a reduced broadening of $\Gamma=5\mu$eV) for finding the chemical potential at low temperatures.
}
and, at this point, we do not attempt to introduce effects of impurities.

\begin{figure}[!t!h]
  \begin{center}
  {\includegraphics[angle=-90,width=.5\textwidth]{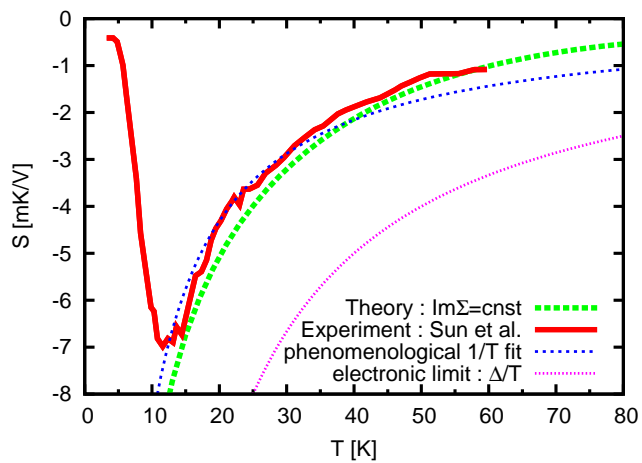}}
      \caption{Thermopower of \feas.
     Shown are  the theoretical Seebeck coefficient for x-polarization, using a constant self-energy, and the GGA band-structure, with the gap scissored to 0.2eV.  
      Experimental results are of Sun \etal \cite{APEX.2.091102} (x-polarization). Also displayed is a simple 
      1/T fit corresponding to the large gap semi-conductor, \eref{eq:Ssc}, yielding $\Delta/2 \delta\lambda-\mu\approx86$meV.
      Further indicated is the largest possible purely electronic Seebeck coefficient for \feas.
}
       \label{figSFeAs2}
      \end{center}
\end{figure}

In \fref{figSFeAs2} we show our theoretical Seebeck coefficient of \feas\ as a
function of temperature (green dashed curve), and compare it to
experimental results\cite{APEX.2.091102} for the same polarization.
The agreement is excellent 
in the intrinsic, i.e.\ not impurity dominated, temperature regime ($T>12$K).

Also shown is a simple fit, using the formula \eref{eq:Ssc} for the large gap semiconductor.
The individual determination of the parameters $\delta\lambda$ and $\mu$ ($\Delta$ given by experiment) is ambiguous, given
the scale of the low temperature thermopower (mV/K) with respect to the high temperature Heikes limit (which is of the order of $k_B/e=86 \mu V/K $). 
In \fref{figSFeAs2} we show results for the large gap model for 
$\Delta/2 \delta\lambda-\mu=85$meV (blue dotted curve), which is compatible with the constraints $\left|\delta\lambda\right|\le 1$, and $\mu\le\Delta/2$.

The decrease of the Seebeck coefficient at low temperature can be understood from our model considerations. This can both be an effect of the scattering rate as well as
the presence of impurities, as seen e.g.\ in \fref{model}(a,b).
In the current {\it ab initio} case, our limited numerical precision (mostly caused by the finite k-mesh) prevents us from endeavoring to include these effects.

\subsection{\fesb}

The situation is entirely different for \fesb.  The maximal measured
Seebeck coefficient $S(T=10\hbox{K})$ is $-45$mV/K.\cite{0295-5075-80-1-17008}  If one takes the maximum
possible asymmetry parameter, $\delta\lambda=1$, and one assumes the chemical
potential to be at the most favorable position, i.e. $\mu=-\Delta/2$, the
charge gap must be larger than $\Delta=0.45$eV to explain the value of
the meassured thermopower in terms of our purely electronic model.
The experimental charge gap, however, is only $\Delta\approx30$meV.
We are thus led to suspect that the large thermopower of \fesb\ at low temperature is not purely of electronic origin.


A possible scenario, mentioned in the literature, is that the very large Seebeck coefficient is mainly caused by a substantial phonon drag, i.e.\ by an electron drift induced by a scattering with phonons.
While there is no conclusive evidence that this effect is operational in \fesb, there are several reasons why it is more likely to be present in this material than in \feas. 
Since the thermopower is a measure for the entropy per carrier, the phonon contribution to the Seebeck coefficient will be proportional to the lattice specific heat times the electron-phonon coupling constant divided by the electron density.
Given Debye temperatures of $348$~K for \fesb\cite{bentien:205105,APEX.2.091102}, and $510$~K for \feas\cite{APEX.2.091102}, the specific heat of \fesb\ will be 
larger than that of \feas.
The charge carrier concentration at temperatures where the thermopower is maximal, on the other hand, is larger for \fesb\cite{sun_dalton}~: for the best
sample $n\sim 8\cdot 10^{14}$/cm$^3$  
whereas for \feas\ $n\sim 5\cdot 10^{14}$/cm$^3$.\cite{sun_dalton}

The electron-phonon coupling is the least accessible ingredient from the theoretical point of view.
Experimentally there are some insinuations~: First of all, the low temperature feature seen in the specific heat of \fesb, that has no analogue in the spin response, and is absent in the arsenide,\footnotemark[\thefnnumber] 
could originate from a substantial electron-phonon coupling, charge ordering, excitonic or polaronic effects from an enhanced coupling to the lattice.

Also the nuclear spin-lattice relaxation rate increases below 40K,\cite{1742-6596-150-4-042040} i.e.\ in the regime where the thermopower starts its huge magnification, whereas an activation law decrease ($\Delta=473$~K) is found above 50K,\cite{1742-6596-150-4-042040} in rough accordance with the intrinsic gap. 

Moreover, optical spectroscopy witnesses a large change in phonon lifetimes across the metal-insulator transition, suggesting an important electron-lattice coupling~\cite{perucci_optics}. 
Recently, also polarized Raman scattering experiments gave indications for a notable electron-phonon coupling, that is strongly temperature dependent below 40K.\cite{PhysRevB.81.144302}
Further, we note that, as expected for substantial phonon drag contributions to the Seebeck coefficient, the magnetothermopower of \fesb\ is very low for those samples that exhibit the largest response without magnetic field~\cite{0295-5075-80-1-17008}. A decrease in the phonon mean free path by non-electronic scattering (i.e.\ in particular by imperfections) is expected to lower the respective effect in the thermopower, and indeed the Seebeck coefficient of polycrystalline samples~\cite{bentien:205105} and thin films~\cite{sun:033710} was found to be significantly smaller than for single crystals, while having the same high temperature behavior~\cite{1742-6596-150-4-042040}. 
Recently, also substituted FeSb$_{2-x}$As$_x $ was investigated~\cite{sun_dalton}.
Interestingly, it was found that the above mentioned increase in the susceptibility, starting at around 50K, is stable with respect to the substitution, whereas 
the shoulder in the resistivity at $10-20$K is flattened out, and the Seebeck coefficient decreases.
In the phonon-drag picture this would, again, be owing to a decrease in the phonon mean free path for non-electronic scattering due to the presence of
the $As$ ``impurities''.

\begin{figure}[!t!h]
  \begin{center}
  {\includegraphics[angle=-90,width=.5\textwidth]{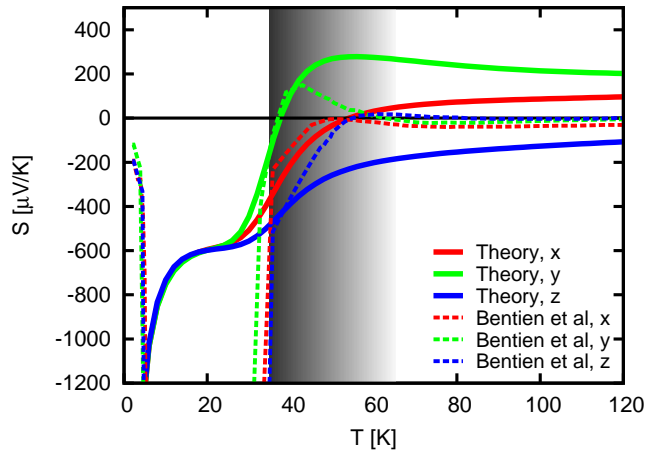}}
      \caption{Thermopower of \fesb.
      Shown are our theoretical together with experimental results from Ref.~\onlinecite{0295-5075-80-1-17008} 
      for measurements along the crystallographic orientations x,y,z, as indicated.
	    We can expect reasonable agreement in the temperature range indicated by the gray area. See text for details.
}
       \label{figSfesb2}
      \end{center}
\end{figure}


Comparing the thermopower of the antimonide and the arsenide (see Fig. 1 in Ref.~\onlinecite{APEX.2.091102}) one notes that
the Seebeck coefficient of \feas\ is larger than that of \fesb\ 
at 35K and higher. This might indicate -- if the phonon drag picture holds -- that the effective electron phonon coupling in \fesb\ has
sufficiently decreased (by umklapp and phonon-phonon scattering) so that the thermopower is now dominated by the electronic degrees of freedom, i.e.\
the larger gap in \feas\ causes a larger response.
Yet, we note that optical spectroscopy\cite{perucci_optics} and some transport measurements\cite{PhysRevB.67.155205,PhysRevB.74.195130} see metallic behavior above 70K, or already above 40K, respectively, an
effect not captured by our one-particle approach. Hence we will focus on the temperature range from 35K upwards to, at best, 70K.

Since the GGA Kohn-Sham spectrum is metallic, we opted for using  
the hybrid functional calculation (see \fref{fig1}(b)), albeit with a gap scissored to the experimental value $\Delta=0.03$eV, to compute the theoretical Seebeck coefficient.
Moreover,
we assume the presence of donor impurities at $E_D=9$meV, corresponding to an activation energy $\delta=\Delta/2-E_D=6$meV as is seen in the resistivity in the range of 5-15K,\cite{0295-5075-80-1-17008} and we use an impurity concentration $n_D=10^{17}$/cm$^3$.
This concentration yields $n_D^++p\approx 7\cdot 10^{16}$/cm$^3$ at 20K, in rough accordance with the respective hole concentration of $4\cdot10^{17}$/cm$^3$ found in Hall measurements~\cite{hu:182108}. 

We again limit the influence of impurities to their effect on the chemical potential.\footnote{To avoid pathologies introduced by the necessary numerical broadening in the spectral function,
we eliminate spurious weight of valence and conduction bands inside the gap.} %
 At high temperature, the latter is linear in $T$ as expected, and, using \eref{muintrinsic},
we find an effective mass ratio $\eta_v/\eta_c=m^*_v/m^*_c=0.23$ when using a constant scattering rate, and the very similar $\eta_v/\eta_c=0.25$ when using the imaginary parts of the self-energy from the GW calculation, i.e. the anisotropy is mainly propelled by the spectral weight and the Fermi velocities, and the GW scattering actually slightly reduces the particle--hole asymmetry in the current case.
We further note that the asymmetry is opposite to that of \feas, where we found $\eta_v/\eta_c=2.5>1$.

Thus obtained Seebeck coefficient is displayed in \fref{figSfesb2}, along with experimental results on single crystals\cite{0295-5075-80-1-17008} 
for  the three polarizations along the crystallographic axes.
In the limited range (discussed aboved), starting at 35K, and extending towards 70K (indicated by the gray gradient in \fref{figSfesb2}), we find good agreement with experiments~:
 Both, the order of magnitudes, as well as the hierarchy of polarizations is captured within our approach.
 Below 35K, the single crystal experiment reaches stellar magnitudes of up to -45mV/K\cite{0295-5075-80-1-17008}, that we argued to be beyond our approach which neglects vertex corrections.  
 Measurements (not shown) using a polycrystalline sample,\cite{bentien:205105} and on films with preponderant $<101>$ orientation\cite{sun:033710}
 display Seebeck coefficients that at low temperatures never surpass -500 and -200$\mu$V/K, respectively, while having the exact same high temperature
 behavior,  
 advocating a disorder or decoherence induced lowering of the electron drift.
 At intermediate temperatures, those experiments agree qualitatively with both the single crystal measurements, and our theoretical results.

\section{conclusions}
\label{conclusions}

In conclusion we have considered the problem of thermoelectricity in
correlated insulators and semiconductors.  We developed a simple toy
model to study how the various many body renormalizations enter the
thermoelectric response.  We used LDA, hybrid density functional
theory and GW methods to carry out a comparative study of two systems
of current experimental and theoretical interest FeAs$_2$ and
FeSb$_2$.
The ratio between strength of the Hubbard $U$ and the bandwidth of
FeAs$_2$ and FeSb$_2$ are comparable and so is the correlation
strength.  In FeAs$_2$ DFT is qualitatively correct, while in FeSb$_2$
correlation effects beyond DFT are essential for obtaining an
insulating ground state, and the one shot GW approximation succeeds in
that respect. Indeed, using this method, we obtained good agreement
with the experimental values of the gap for both materials.

The tools developed in this work were sufficient to describe the
thermoelectric response of FeAs$_2$ quantitatively. This framework is
not as successful for the FeSb$_2$ compound, and in particular it
fails to explain the remarkably high low temperature thermopower
discovered by Bentien \etal\cite{0295-5075-80-1-17008}. Our work implies that the
latter cannot be understood in the context of local correlations, and
one should focus either on vertex corrections to the transport
coefficients, or on non local self energy effects characteristic to
the proximity to a quantum critical point.
In this context we notice that within LDA+U this material is close to
a ferromagnetic instability~\cite{luko_fesb2}.

An important form of vertex corrections describe the phonon drag
effect. A framework to estimate quantitatively
these effects in conjunction with {\it ab inito} methods, are currently not
available.  Above, we mentioned several experimental findings that 
suggest the presence of this mechanism in \fesb, providing a strong incentive to
further development in this vein.

Future work should include explicit calculations on correlated
insulators using LDA+DMFT to compare with the results of the toy model
calculations.  Furthermore, the investigation of vertex corrections on
the thermoelectricity together with the effects of non local self
energies that go beyond the quasiparticle approximation should be
considered.

\section*{Acknowledgments}

We thank K. Behnia, and J. Snyder for interesting discussions.
This international collaboration is supported by the NSF-materials
world network under grant number NSF DMR 0806937 and by the PUF
program.  We are grateful to the KITP who hosted the program "Towards
Materials Design Using Strongly Correlated Electron Systems" where
part of this research was carried out. 
JMT further thanks AIST, Japan, for hospitality.


\end{document}